\documentclass[11pt, a4paper, copyright, goog]{google}

\usepackage[authoryear, sort&compress, round]{natbib}
\usepackage{bbm}
\usepackage{adjustbox}
\usepackage{makecell}
\usepackage{multirow}
\usepackage[table]{xcolor}
\usepackage{wrapfig}
\usepackage{colortbl}
\definecolor{indomain}{rgb}{0.8, 0.8, 1.0}
\definecolor{ood}{rgb}{1.0, 0.902, 0.8}
\usepackage{tcolorbox}

\newtcolorbox{prompt}[1]{
    enhanced,
    drop shadow=black!5!white,
    left=4mm,
    right=4mm,
    top=2mm,
    bottom=2mm,
    boxsep=0mm,
    rounded corners,
    title=#1,    fontupper=\footnotesize\linespread{0.9}\fontfamily{lmr}\selectfont,
    }
    
\title{
Bridging Textual Profiles and Latent User Embeddings for Personalization
}
\newcommand{\ourmethod}{\textsc{BLUE}}

\correspondingauthor{ztan3@nd.edu, xiangzhai@google.com}

\reportnumber{0001} % Leave blank if n/a

\author[1]{Zhaoxuan Tan}
\author[2]{Xiang Zhai}
\author[2]{Yan Zhu}
\author[1]{Meng Jiang}
\author[2]{Mohamed Hammad}

\affil[1]{University of Notre Dame}
\affil[2]{Google}
\begin{abstract}
Personalized systems rely on user representations to connect behavioral history with downstream recommendation applications. Existing methods typically employ either supervised latent user embeddings, which are effective for retrieval but difficult to interpret, or textual user profiles, which are interpretable but challenging to optimize for downstream utility due to lack of direct supervision. To bridge this gap, we present \ourmethod{}, a reinforcement learning framework that unifies these two forms of user representation by aligning language-based user profiles with embedding-based recommendation objectives. Given a user’s interaction history, \ourmethod{} leverages a profiler Large Language Model (LLM) to generate textual profiles, while an embedding model provides reward signals. This encourages the resulting textual representations to move closer to positive items and farther from negative ones in the embedding space. We further introduce a text-space supervision signal based on next-item prediction, ensuring the learned profiles remain both semantically meaningful and highly effective for downstream retrieval. Experiments on Amazon Reviews 2023 and Google Local Reviews in zero-shot sequential recommendation settings demonstrate that \ourmethod{} consistently outperforms strong baselines under both frozen and trainable embedding conditions. Notably, \ourmethod{} achieves clear gains in cross-domain transfer, highlighting the strong generalization ability of the learned user profiles. Furthermore, these generated profiles provide superior personalized context for question answering compared to raw user histories or alternative profile optimization methods. Overall, these results show that \ourmethod{} provides an effective way to unify interpretable textual profiling with discriminative latent embeddings for personalization. 
\end{abstract}

\begin{document}

\maketitle

\section{Introduction}

User profiling is a core component of personalized systems because it connects raw user behavior with downstream decisions about what content, products, or responses should be presented to each user \citep{tan2023user, zhao2024recommender}. By adapting to individual preferences and behavioral patterns, personalization can improve user experience across many applications, including recommendation \citep{bang2025llm, wang2025lettingo,chen2025memory}, user simulation \citep{yoon2024evaluating}, personalized chatbots \citep{cheng2024dialogues,DBLP:conf/acl/Tan0LWY0C000NY025, tan2025instant,tan2024democratizing}, healthcare \citep{zhang2024llm}, education \citep{xu2024large}, and more.

A central design choice in personalization is how to represent the user. Existing methods mainly rely on two forms of user representation: textual user profiles and latent user embeddings. Textual user profiles describe user preferences in natural language \citep{zhou2024language}, while latent user embeddings map users into dense vectors learned from sequential interaction histories, often with dual-encoder architectures \citep{yang2024unifying} or sequential recommendation models \citep{fang2020deep,li2024calrec}. In practice, latent embeddings are the dominant choice in recommendation systems because they are efficient for retrieval and often achieve strong ranking performance \citep{zhao2023embedding}. However, latent embeddings have limited interpretability. They compress complex and noisy interaction histories into a single dense vector, which makes it hard to distinguish high-level preferences from behavioral noise. As a result, they often act as a black box: strong for ranking, but offering little insight into what the model has learned about the user or why a recommendation is made \citep{zhou2024language,ramos2024transparent}. This compression can also lose semantic information about long-term preferences and behavioral patterns. In contrast, language-based user profiles offer a more interpretable representation in text space. With the help of large language models (LLMs), a system can summarize user interests and behavioral patterns into concise and human-readable descriptions \citep{wang2025lettingo,zhou2024language}. Such profiles are easier to inspect, can capture richer semantics, and can support reasoning over sparse or noisy histories. However, existing profile generation methods still face a key challenge. Most rely on manually crafted prompts and predefined profile formats, which may not align well with downstream personalization tasks. More importantly, training an LLM to generate high-quality user profiles is difficult because there is often no single ground-truth profile for a user \citep{rahimzadeh2025millions}. Unlike embedding learning in sequential recommendation, where the next interacted item provides a clear supervision signal, user preferences, habits, and intent in user profiles can be summarized in many valid ways in natural language \citep{wu2025rlpf}.

\begin{wrapfigure}{r}{0.5\textwidth}

    \centering
    \includegraphics[width=0.96\linewidth]{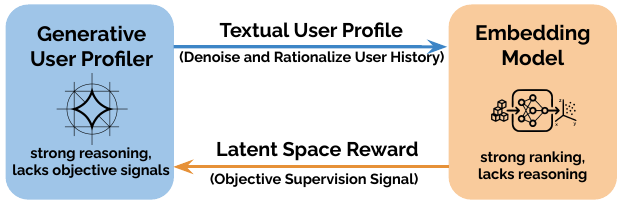}
    \caption{\ourmethod{} bridges textual user profiles and latent user embeddings for personalization. A generative profiler produces an interpretable user profile from raw history, while an embedding model provides latent-space reward signals to optimize profile generation for retrieval.}
    \label{fig:teaser}
\end{wrapfigure}

This creates a gap between the two dominant forms of user representation. Latent embeddings are strong for retrieval but weak in interpretability, while textual profiles are interpretable but difficult to optimize for downstream utility. In this work, we aim to bridge these two representations. We propose to \textbf{B}ridge textua\textbf{L} profiles and latent \textbf{U}ser \textbf{E}mbeddings for personalization (\ourmethod{}), a reinforcement learning framework that connects the semantic transparency of language-based user profiles with the discriminative ranking ability of latent user embeddings. In \ourmethod{}, the textual profile serves as a semantic summary and rationale for user behavior, while the embedding model provides a supervision signal that guides profile generation toward representations that are useful for recommendation (Figure~\ref{fig:teaser}). In this way, the profile is not optimized only to sound plausible in natural language, but also to improve downstream ranking quality in embedding space.

Specifically, given a raw user behavior sequence, we prompt a profiler LLM to generate multiple candidate user profiles. We then combine the generated profile with the user history and feed them into an embedding model to obtain a user representation for ranking against item embeddings. The embedding model is kept frozen while the profiler LLM is optimized with Group Relative Policy Optimization (GRPO) \citep{shao2024deepseekmath}. The reward is defined from contrastive signals in the latent space, encouraging generated profiles whose induced user embeddings are closer to the positive item and farther from negative items. We further include a text-space supervision signal based on next-item multiple-choice prediction, where the generated user profile is fed back into the profiler LLM and perform multiple choice on the next item prediction task, and obtain binary correctness reward. As a result, the profiler learns to generate user profiles that are both semantically meaningful in text space and useful for retrieval in embedding space. These generated profiles can be used as concise personalized context for downstream personalized question answering, and they can also improve recommendation performance under both frozen and fine-tuned embedding settings.

% We evaluate our framework on Amazon Reviews 2023 \citep{hou2024bridging} and Google Local Reviews \citep{li2022uctopic} data under the zero-shot sequential recommendation setting \citep{li2025llm}. Specifically, we train the profiler model on the Amazon Clothing\_Shoes\_and\_Jewelry domain and test on out-of-domain Books, Electronics, Sports, and Google Reviews data. Our results show that \ourmethod{} consistently outperforms strong baselines while keeping the embedding model frozen. Furthermore, by further finetuning the embedding model, \ourmethod{} performance can further improve while keeping the advantage over baselines. We also show that the generated textual user profiles provide better personalized context for question answering than raw history or alternative profile optimization methods. Extensive experiments also present in-domain performance across domains and the strong pareto optimal tradeoff between performance and efficiency and 

We evaluate our framework on Amazon Reviews 2023 \citep{hou2024bridging} and Google Local Reviews \citep{li2022uctopic} data under the zero-shot sequential recommendation setting \citep{li2025llm}. Specifically, we train the profiler model on the Amazon Clothing\_Shoes\_and\_Jewelry domain and test on out-of-domain Books, Electronics, Sports, and Google Reviews data. Our results show that \ourmethod{} consistently outperforms strong baselines while keeping the embedding model frozen. Furthermore, by further finetuning the embedding model, \ourmethod{} can achieve additional gains while still maintaining its advantage over competing methods. We also show that the generated textual user profiles provide better personalized context for question answering than raw history or alternative profile optimization methods. Extensive experiments further demonstrate strong in-domain performance across datasets, a favorable Pareto trade-off between effectiveness and efficiency, and improved latent user structure in both quantitative analysis and qualitative case studies.

To summarize, our contributions are twofold. First, we propose \ourmethod{}, a reinforcement learning framework that jointly connects textual user profiling and latent embedding optimization for personalization, showing that latent-space contrastive reward can serve as an effective supervision signal for training a profiler LLM without requiring ground-truth user summaries. Second, we demonstrate that the resulting profiles support two complementary personalization goals: improving retrieval quality for sequential recommendation and providing high-quality, interpretable user context for personalized question answering.

\section{Methodology}

We aim to bridge latent user embeddings and textual user profiles by connecting them during training. Instead of treating profile generation as a standalone language task, we model it as a policy that summarizes user history into a concise natural language profile, which serves both as an interpretable description and as input to an embedding model for recommendation. The key idea is to use the downstream utility of the induced embedding as supervision: rather than matching a ground-truth summary, the profiler is optimized to generate profiles that improve downstream prediction. In this pipeline, a user’s interaction history is first converted into a textual profile, which is then encoded into a latent representation for retrieval, and the resulting performance signal is fed back to refine the profile generation. This allows the learned profiles to remain semantically meaningful while also capturing preference information that is useful for personalization.

\subsection{Problem Formulation}

Let $\mathcal{U}$ and $\mathcal{V}$ denote the sets of users and items, respectively. For a given user $u \in \mathcal{U}$ at time $t$, let $\mathcal{H}_u = \{h_u\}^{\leqslant t}$ denote the user's historical interaction sequence before time $t$. Each interaction is represented as a tuple $h_u = (v, s)$, where $v \in \mathcal{V}$ is the interacted item and $s$ is the associated score or rating. The item interacted with after time $t$ is treated as the target positive item and denoted by $v_{pos}$. Our goal is to learn a profiler model $\pi_{\theta}$ that maps raw user history to a textual user profile,
\begin{align*}
    p_u = \pi_{\theta}(\mathcal{H}_u),
\end{align*}
such that the user profile serves two roles: \textit{(1)} an interpretable summary of user preferences for downstream language tasks, and \textit{(2)} an effective representation for sequential recommendation when encoded by an embedding model. Instead of relying on ground-truth summaries, we optimize $\pi{\theta}$ using downstream signals that capture both text-space and embedding-space usefulness.

\begin{figure}
    \centering
    \includegraphics[width=0.9\linewidth]{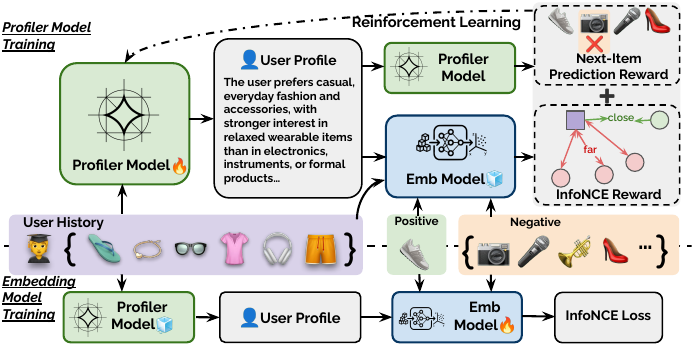}
    % \caption{Overview of \ourmethod{}. Given a user history, a profiler generates a textual user profile, optimized with reinforcement learning using two signals: an embedding-space contrastive reward from a frozen embedding model and a text-space next-item prediction reward. This enables \ourmethod{} to learn profiles that are both interpretable and useful for sequential recommendation and question answering. The embedding model can be further finetuned on user behavior by using generated user profile as input}
    \caption{Overview of \ourmethod{}. Given a user history, a profiler generates a textual user profile and is optimized with reinforcement learning using a text-space next-item prediction reward and an embedding-space contrastive reward from a frozen embedding model. The learned profile can then be used to train the embedding model with InfoNCE loss, yielding user representations that are both interpretable and effective for sequential recommendation and personalized question answering.}
    \label{fig:overview}
\end{figure}

\subsection{Bridging Texutal User Profile and Latent User Embeddings (\ourmethod{})}

Figure~\ref{fig:overview} illustrates the overall framework of \ourmethod{}. Starting from the raw user behavior sequence $\mathcal{H}_u$, we prompt a profiler LLM to generate a textual profile $p_u$ that summarizes the user's interests, preferences, and behavioral patterns:
\begin{align*}
    p_u \sim \pi_{\theta}( \cdot \mid \mathcal{P}_{\mathrm{Gen}}(\mathcal{H}_u)).
\end{align*}

This generated profile is not treated as the final output by itself. Instead, it functions as a bridge between the user's raw behavioral history and two downstream personalization spaces. In the text space, it provides an explicit natural language summary that can support personalized question answering. In the latent space, it is encoded into user representations that can be used for item retrieval and ranking. To preserve both short-term behavioral evidence and high-level preference abstraction, we construct two user embedding variants using a frozen embedding model $\mathrm{Emb}_{\phi}(\cdot)$:
\begin{align*}
    e_u^{(1)} = \mathrm{Emb}_{\phi}(\mathcal{H}_u' \; || \; p_u), \; \; \;
    e_u^{(2)} = \mathrm{Emb}_{\phi}(p_u).
\end{align*}
Here, $e_u^{(1)}$ combines recent behavioral context $\mathcal{H}_u'$ (the most recent 10 interactions) with the generated profile, while $e_u^{(2)}$ reflects how much useful preference information is captured by the profile alone. The profile encodes long-term user preferences and behavior patterns, while the recent history captures short-term dynamics.

\paragraph{Embedding-Space Reward.}
To encourage the generated profile to induce a user representation that is effective for sequential recommendation, we define an embedding-space reward based on bidirectional contrastive alignment between the user representation and the target item embeddings. For each embedding variant $e_u^{(k)}$, where $k \in {1,2}$, we consider the average of user-to-item and item-to-user contrastive objectives under an InfoNCE-style formulation \citep{Oord2018RepresentationLW}. The user-to-item term pulls the user embedding closer to the positive item while pushing it away from negative items, whereas the item-to-user term separates the target item from other users in the batch. Assuming a symmetric similarity function $\mathrm{sim}(\cdot,\cdot)$, instantiated as cosine similarity, the objective is
\begin{align*}
R_{\mathrm{avg}}(e_u^{(k)}) =
\frac{1}{2}\Big(R_{U \to I}(e_u^{(k)}) + R_{I \to U}(e_u^{(k)})\Big)
=
\log \frac{\exp(\mathrm{sim}(e_u^{(k)}, h_{v^+}) / \tau)}
{\sqrt{Z_U(e_u^{(k)}) \cdot Z_I(e_u^{(k)})}}.
\end{align*}
where the $h_{v^+}$ is the positive next interacted item of user $u$, $\mathcal{V}^-$ is the set of negative items, $\tau$ is the temperature for contrastive reward, $\mathcal{U}^-$ denotes all other users in batch. The corresponding partition functions $Z_U(e_u^{(k)})$ and $Z_I(e_u^{(k)})$ are given by
\begin{align*}
Z_U(e_u^{(k)}) &= \exp(\mathrm{sim}(e_u^{(k)}, h_{v^+}) / \tau)
+ \sum_{j \in \mathcal{V}^-} \exp(\mathrm{sim}(e_u^{(k)}, h_j) / \tau), \\
Z_I(e_u^{(k)}) &= \exp(\mathrm{sim}(h_{v^+}, e_u^{(k)}) / \tau)
+ \sum_{j \in \mathcal{U}^{-}} \exp(\mathrm{sim}(h_{v^+}, e_j^{(k)}) / \tau).
\end{align*}
This reward encourages the generated profile to align the user representation with the ground-truth next item while separating it from negative items and users. We then aggregate the rewards from the two representation variants:
\begin{align*}
R_{\mathrm{emb}}^u = R_{\mathrm{avg}}(e_u^{(1)}) + R_{\mathrm{avg}}(e_u^{(2)}).
\end{align*}
This term provides supervision from the latent space, guiding the textual profile to improve downstream retrieval performance rather than only optimizing for surface-level fluency.

\paragraph{Text-Space Reward.}
While the embedding-space reward improves retrieval quality, it does not explicitly ensure that the generated profile is informative and usable in natural language. To maintain the semantic quality of the profile, we introduce a text-space reward that directly evaluates whether the profile contains sufficient information to support language-based reasoning about user behavior.

Concretely, we formulate next-item prediction as a multiple-choice question answering task. Given the generated profile $p_u$ as context, we construct a prompt $\mathcal{P}_{\mathrm{pred}}(p_u)$ that asks the model to select the next interacted item from a candidate set. The profiler model $\pi_{\theta}$ then produces a prediction:
\begin{align*}
    \hat{y}_u \sim \pi_{\theta}(\cdot \mid \mathcal{P}_{\mathrm{pred}}(p_u)).
\end{align*}
where $\hat{y}_u$ denotes the predicted item. The ground-truth label is the actual next item in the sequence, denoted by $y_u = v_u^+$.
\begin{align*}
    R_{\mathrm{text}}^u = \mathbbm{1}(\hat{y}_u = y_u), \quad \text{where} \quad y_u = v_u^+.
\end{align*}

This reward encourages the generated profile to contain explicit and usable preference information that supports language-based reasoning. In contrast to the embedding-space reward, which operates in a latent space, this term directly evaluates whether the profile alone provides sufficient information to infer the user’s next action in natural language. As a result, it helps maintain the interpretability and practical usefulness of the generated profiles for downstream LLM-based personalization tasks.

\paragraph{Final reward and policy optimization.}
We combine the embedding-space and text-space signals into a unified reward for each sampled profile. For a given user $u$ and a sampled profile $p_u^{(i)}$ from the current policy $\pi_\theta$, the total reward is defined as
\begin{align*}
    R_{\mathrm{final}}^u(p_u^{(i)}) = R_{\mathrm{text}}^u(p_u^{(i)}) + R_{\mathrm{emb}}^u(p_u^{(i)}),
\end{align*}
where $R_{\mathrm{text}}^u$ evaluates next-item prediction accuracy in text space, and $R_{\mathrm{emb}}^u$ measures contrastive alignment quality in embedding space. To optimize the profiler model $\pi_\theta$, we adopt Group Relative Policy Optimization (GRPO). For each user $u$, we sample a group of $G$ candidate profiles $\{p_u^{(i)}\}_{i=1}^G$ from the current policy. The objective is defined as
\begin{align*}
\mathcal{J}_{\mathrm{GRPO}}(\theta)
=
\mathbb{E}_{u}
\left[
\frac{1}{G}\sum_{i=1}^{G}
\left(
\min\!\big(
r_{u,i}\hat{A}_{u,i},
\mathrm{clip}(r_{u,i},1-\epsilon,1+\epsilon)\hat{A}_{u,i}
\big)
-\beta\,\mathbb{D}_{\mathrm{KL}}(\pi_\theta \,\|\, \pi_{\mathrm{ref}})
\right)
\right],
\end{align*}
where $r_{u,i}$ is the probability ratio defined as
\begin{align*}
    r_{u,i} = \frac{\pi_\theta(p_u^{(i)} \mid \mathcal{P}_{\mathrm{Gen}}(\mathcal{H}_u))}{\pi_{\mathrm{ref}}(p_u^{(i)} \mid \mathcal{P}_{\mathrm{Gen}}(\mathcal{H}_u))},
\end{align*}
which compares the likelihood of the sampled profile under the current policy $\pi_\theta$ and a reference policy $\pi_{\mathrm{ref}}$. The clipping parameter $\epsilon$ controls the trust region of policy updates, and $\beta$ balances the strength of the Kullback–Leibler (KL) regularization term $\mathbb{D}_{\mathrm{KL}}(\pi_\theta \,\|\, \pi_{\mathrm{ref}})$, which stabilizes training by preventing large deviations from the reference policy. The normalized advantage $\hat{A}_{u,i}$ is computed within each group as
\begin{align*}
    \hat{A}_{u,i} = \frac{R_{\mathrm{final}}^u(p_u^{(i)}) - \mu_u}{\sigma_u + \epsilon},
\end{align*}
where $\mu_u = \frac{1}{G}\sum_{i=1}^G R_{\mathrm{final}}^u(p_u^{(i)})$ and $\sigma_u$ is the corresponding standard deviation across the $G$ sampled profiles for user $u$. This group-wise normalization reduces variance and enables stable comparison among candidate profiles generated from the same user history. Overall, this formulation does not assume a single correct textual profile for each user. Instead, profiles are optimized based on their functional quality: a profile is preferred if it yields higher reward by being both informative in text space and effective in embedding space. By jointly optimizing these signals, \ourmethod{} learns profiles that support both interpretable reasoning and accurate personalized retrieval in text and latent space.

\subsection{Embedding Model Optimization}

% We consider two sequential recommendation settings. In the first setting, the embedding model is kept frozen, which reflects realistic scenarios where the embedding service is closed-source or cannot be updated. In the second setting, the embedding model is further finetuned using supervised contrastive learning \citep{khosla2020supervised}. This allows us to study whether the learned textual profiles remain useful not only as fixed prompts for retrieval, but also as improved supervision signals when adapting the embedding model itself.

The embedding model in \ourmethod{} primarily serves as a source of supervision for guiding profile generation. In many practical scenarios, the embedding model is a fixed component, such as a deployed retrieval system or a closed-source service, and cannot be updated. In this setting, our method already improves personalization by optimizing textual profiles to better align with the existing embedding space. However, the learned textual profiles can also be viewed as structured and denoised summaries of user behavior, which may provide higher-quality supervision signals than raw interaction histories. This motivates a second setting where the embedding model is further finetuned using the generated profiles, while the profiler model is frozen. Importantly, this step is optional, but it allows us to study whether improved profiles can also enhance the embedding space training.

Specifically, we consider supervised contrastive learning \citep{khosla2020supervised} to adapt the embedding model. Given a mini-batch of $B$ user-item pairs, let $e_u$ denote the user embedding derived from the profile-augmented context, and let $h_{u,v^+}$ denote the embedding of the corresponding ground-truth next item. The user-to-item and item-to-user objectives are:
\begin{align*}
    \mathcal{L}_{U \to I} =
    - \frac{1}{B} \sum_{u=1}^{B}
    \log
    \frac{\exp(\mathrm{sim}(e_u, h_{u,v^+}) / \tau)}
    {\sum_{j=1}^{B} \exp(\mathrm{sim}(e_u, h_{j,v^+}) / \tau)},\;\;\mathcal{L}_{I \to U} =
    - \frac{1}{B} \sum_{u=1}^{B}
    \log
    \frac{\exp(\mathrm{sim}(h_{u,v^+}, e_u) / \tau)}
    {\sum_{j=1}^{B} \exp(\mathrm{sim}(h_{u,v^+}, e_j) / \tau)}.
\end{align*}

The overall embedding optimization objective is
\begin{align*}
    \mathcal{L}_{\mathrm{emb}} = \mathcal{L}_{U \to I} + \mathcal{L}_{I \to U}.
\end{align*}

By minimizing $\mathcal{L}_{\mathrm{emb}}$, we adapt the embedding model so that user profiles and item representations lie in a more discriminative shared space. This setting allows us to evaluate whether the generated textual profiles not only improve retrieval as fixed inputs, but also act as effective supervision signals for learning better embedding functions.

\section{Experimental Settings}

\subsection{Models and Datasets}

\begin{wraptable}{R}{0.5\textwidth}
% \begin{table}[t]
    \caption{Dataset statistics used in our experiments. We train on Amazon Clothing and evaluate cross-domain generalization on Amazon Books, Amazon Electronics, Amazon Sports, and Google Reviews.}
    \centering
    \begin{adjustbox}{max width=1\linewidth}
        \begin{tabular}{llrrrr}
        \toprule[1.5pt]
        \textbf{Dataset} & \textbf{Split} & \textbf{\# Items} & \textbf{\# Users} & \textbf{Avg. Seq Len} & \textbf{\# Examples} \\
        \midrule[0.75pt]
        \multirow{2}{*}{\makecell[l]{\textsc{Amazon Clothing}}} & Train & 983,565 & 79,897 & 22.10 & 106,946 \\
        & Test & 131,693 & 1,209 & 22.90 & 1,209 \\
        \midrule
        \multirow{2}{*}{\textsc{Amazon Books}} & Train & 746,516 & 31,060 & 26.66 & 44,420 \\
        & Test & 141,357 & 1,252 & 31.39 & 1,252 \\
        \midrule
        \multirow{2}{*}{\textsc{Amazon Electronics}} & Train & 534,876 & 42,310 & 25.52 & 58,422 \\
        & Test & 126,392 & 1,256 & 27.30 & 1,256 \\
        \midrule
        \multirow{2}{*}{\makecell[l]{\textsc{Amazon Sports} \\\textsc{and Outdoors}}} & Train & 205,036 & 14,529 & 22.98 & 22,673 \\
        & Test & 98,911 & 1,295 & 24.06 & 1,295 \\
        \midrule
        \multirow{2}{*}{\textsc{Google Reviews}} & Train & 420,373 & 72,273 & 67.19 & 100,000 \\
        & Test & 136,037 & 1,338 & 74.90 & 1,338 \\
        \bottomrule[1.5pt]
        \end{tabular}%
    \end{adjustbox}
    \vspace{-0.3cm}
    \label{tab:data_stats}
% \end{table}
\end{wraptable}
We instantiate \ourmethod{} using \textit{Gemma3-1B-it} \citep{DBLP:journals/corr/abs-2503-19786} as the profiler model and \textit{Qwen3-Embedding-0.6B} \citep{zhang2025qwen3} as the embedding model. We conduct experiments on two real-world datasets: Amazon Reviews 2023 \citep{hou2024bridging} and Google Local Reviews \citep{li2022uctopic}, following the zero-shot sequential recommendation setting in \citet{li2025llm}. Specifically, we train on the Amazon Clothing\_Shoes\_and\_Jewelry domain and evaluate cross-domain generalization on Amazon Books, Electronics, Sports, and Google Reviews within New York state. To ensure sufficient behavioral context, we retain only users with at least 20 and at most 1,000 historical interactions. For each user, we define the target item as the most recent interaction within a one-hour window of the final timestamp, and treat all preceding interactions as the user history. During evaluation, we construct a candidate set consisting of the ground-truth item and 99 randomly sampled negative items that the user has not interacted with. The model is then evaluated based on its ability to rank the target item among these candidates. For personalized question answering, we ask the model to perform user's next item prediction and user rating prediction tasks by using user profile as context. Dataset statistics are summarized in Table~\ref{tab:data_stats}.

\subsection{Baselines}

\paragraph{Embedding-Frozen Sequential Recommendation.}
We first consider the setting where the embedding model is frozen, which is common for proprietary or off-the-shelf embedding services that cannot be further adapted. We compare \ourmethod{} against three groups of baselines. The first group consists of \textit{raw-history baselines}, including \textit{All History} and \textit{10 History}, which use the full user history or the most recent 10 interactions as input. The second group consists of \textit{profile-only baselines}, including \textit{Base Profile}, \textit{RLAIF}~\citep{rlaif}, \textit{RLPF}~\citep{wu2025rlpf}, and \textit{LangPTune}~\citep{langptune}. The third group consists of \textit{hybrid baselines} that combine the most recent 10 interactions with a generated profile. These comparisons test whether optimizing textual profiles can improve retrieval beyond directly using raw behavioral history.

\paragraph{Embedding-Adaptive Sequential Recommendation.}
We next study the setting where the embedding model is further trained and evaluate cross-domain transfer by training on the Amazon Clothing domain and directly testing on other domains without target-domain fine-tuning. We compare against standard sequential recommendation baselines that take LLM embeddings as input, including \textit{SASRec}~\citep{kang2018self}, \textit{BERT4Rec}~\citep{sun2019bert4rec}, and \textit{UniSRec}~\citep{hou2022towards}, where the \textsc{-sem} denotes using semantic embeddings produced by LLMs to enable domain transfer. We also compare with profile-based supervised contrastive baselines that train the embedding model using different forms of user context, including \textit{10H+SCL}, \textit{Base+SCL}, \textit{RLAIF+SCL}, \textit{RLPF+SCL}, and \textit{LangPTune+SCL}, where the context is constructed by concatenating the most recent 10 interactions and generated user profiles and \textsc{SCL} denotes training embedding model with supervised contrastive learning. This setting evaluates whether the learned profiles capture transferable user preferences under domain shift.

\paragraph{Personalized Question Answering.}
To test whether the generated profiles are useful as natural language user context, we evaluate personalized question answering on two tasks: next-item multiple-choice prediction and user rating prediction. We compare against profile-only baselines (\textit{Base Profile}, \textit{RLAIF}, \textit{RLPF}, and \textit{LangPTune}), raw-history baselines (\textit{All History} and \textit{10 History}), and hybrid baselines that combine recent interaction history with generated profiles. All question answering tasks are performed by Gemma3-12B-it model with decoding temperature set to 0. This evaluation complements the recommendation results by measuring whether the profiles are concise, informative, and useful for downstream language-based personalization application.

\paragraph{Evaluation.}
For both embedding-frozen and embedding-adaptive sequential recommendation, we report NDCG@10 (\textit{N@10}) and Recall@10 (\textit{R@10}). For personalized question answering, we use accuracy for next-item prediction and mean absolute error (MAE) for user rating prediction. Lower MAE indicates better performance, while higher values are better for all other metrics.

\begin{table}[t]
   % \vspace{-0.3cm}
   % \caption{Experimental results under the embedding-frozen sequential recommendation. The embedding model is kept fixed, while different user context construction methods are used for ranking. \textit{N@10} and \textit{R@10} denote NDCG@10 and Recall@10, respectively, and higher values indicate better performance. \colorbox{indomain}{Blue} denotes the source-domain Amazon Clothing benchmark, while \colorbox{ood}{orange} denotes out-of-domain test benchmarks without further notice. }
      % \caption{Experimental results under the embedding-frozen sequential recommendation. The embedding model is kept fixed, while different user context construction methods are used for ranking. \textit{N@10} and \textit{R@10} denote NDCG@10 and Recall@10, respectively, and higher values indicate better performance. \textbf{Bold} indicates the best result, and \underline{underlined} indicates the second-best result. \colorbox{indomain}{Blue} denotes the source-domain Amazon Clothing benchmark, while \colorbox{ood}{orange} denotes out-of-domain test benchmarks without further notice. }
  \caption{Embedding-frozen sequential recommendation results. The embedding model is fixed, and only the user context differs. \textit{N@10} and \textit{R@10} denote NDCG@10 and Recall@10. Higher is better. \textbf{Bold} and \underline{underlined} indicate the best and second-best results. \colorbox{indomain}{Blue} denotes the source-domain Amazon Clothing benchmark, and \colorbox{ood}{orange} denotes out-of-domain benchmarks. Throughout all result tables, the best results are marked with *($\tau$-test, $p\leqslant$0.05) compared with the best baselines, unless otherwise specified.}
   \centering
   \small
   \begin{adjustbox}{max width=1\linewidth}
   \begin{tabular}{lcc|cccccccc}
   \toprule[1.2pt]
   \multirow{2}{*}{\textbf{Method}} & \multicolumn{2}{c}{\cellcolor{blue!20}\textbf{Amazon Clothing}} & \multicolumn{2}{c}{\cellcolor{orange!20}\textbf{Amazon Books}} & \multicolumn{2}{c}{\cellcolor{orange!20}\textbf{Amazon Electronics}} & \multicolumn{2}{c}{\cellcolor{orange!20}\textbf{Amazon Sports}} & \multicolumn{2}{c}{\cellcolor{orange!20}\textbf{Google Reviews}} \\
   \cmidrule(lr){2-3} \cmidrule(lr){4-5} \cmidrule(lr){6-7} \cmidrule(lr){8-9} \cmidrule(lr){10-11}
   & \textit{N@10} & \textit{R@10} & \textit{N@10} & \textit{R@10} & \textit{N@10} & \textit{R@10}& \textit{N@10} & \textit{R@10}& \textit{N@10} & \textit{R@10} \\
   \midrule[0.75pt]
   % \multicolumn{11}{c}{ \cellcolor{gray!20}{\textit{Qwen3-Embedding-0.6B}}}\\
   \textsc{All History} & 0.091 & 0.175 & 0.134 & 0.256 & 0.105 & 0.209 & 0.131 & 0.254 & 0.123 & 0.242 \\
   \textsc{10 History} & 0.085 & 0.168 & 0.109 & 0.215 & 0.086 & 0.177 & 0.125 & 0.246 & \underline{0.197} & 0.342 \\
   \textsc{Base Profile} & 0.107 & 0.206 & 0.228 & 0.398 & 0.094 & 0.198 & 0.161 & 0.318 & 0.188 & 0.335 \\
   \textsc{RLAIF Profile} & 0.126 & 0.246 & \underline{0.245} & \underline{0.412} & \underline{0.123} & \underline{0.237} & 0.176 & 0.331 & 0.194 & \underline{0.352} \\
   \textsc{RLPF Profile} & 0.116 & 0.219 & 0.238 & 0.383 & 0.122 & 0.228 & 0.172 & 0.323 & 0.181 & 0.333 \\
   \textsc{LangPTune Profile} & \underline{0.140} & \underline{0.253} & 0.231 & 0.403 & 0.110 & 0.217 & \underline{0.177} & \underline{0.341} & 0.165 & 0.301 \\
   \rowcolor{gray!20}\textsc{\ourmethod{} Profile} (Ours) &  \textbf{0.143} & \textbf{0.266$^*$} & \textbf{0.291$^*$} & \textbf{0.466$^*$} & \textbf{0.130$^*$} & \textbf{0.250$^*$} &  \textbf{0.196$^*$} & \textbf{0.368$^*$} & \textbf{0.202} & \textbf{0.367$^*$} \\
   \hline
   \textsc{10 H+Base Profile} & 0.137 & 0.255 & 0.277 & 0.442 & 0.114 & 0.237 & 0.180 & 0.345 & 0.226 & 0.387\\
   \textsc{10 H+RLAIF Profile} & 0.135 & 0.252 & 0.252 & 0.412 & 0.128 & \underline{0.250} &  0.180 & 0.208 & 0.231 & 0.393\\
   \textsc{10 H+RLPF Profile} & 0.128 & 0.244  & 0.244 & 0.401 & \underline{0.129} & 0.246 & 0.177 & 0.331 & 0.234 & 0.394 \\
   \textsc{10 H+LangPTune Profile} & \underline{0.153} & \underline{0.269} & \underline{0.279} & \underline{0.453} & 0.123 & 0.247 &  \underline{0.187} & \underline{0.363} &  \underline{0.250} & \underline{0.415} \\
   \rowcolor{gray!20}\textsc{10 H+\ourmethod{} Profile} (Ours) & \textbf{0.173$^*$} & \textbf{0.306$^*$} & \textbf{0.329$^*$} & \textbf{0.516$^*$} & \textbf{0.143$^*$} & \textbf{0.267$^*$} & \textbf{0.198$^*$} & \textbf{0.376$^*$} & \textbf{0.264} & \textbf{0.436$^*$}\\
  
   \bottomrule[1.2pt]
   \end{tabular}
   \end{adjustbox}
   \vspace{-0.3cm}
   \label{tab:frozen_emb}
\end{table}
 
\section{Results}

\paragraph{Embedding-Frozen Sequential Recommendation.}
Table~\ref{tab:frozen_emb} reports retrieval results when the embedding model is fixed and only the user context formulation differs. Three patterns stand out. First, learned textual profiles are generally more effective than raw interaction history alone. Compared with \textit{All History} and \textit{10 History}, profile-based methods achieve clearly stronger performance on most domains, suggesting that compressing user behavior into a concise semantic profile can provide a more useful retrieval signal than directly encoding long or recent histories. Among profile-only methods, \textit{\ourmethod{} Profile} performs best overall, achieving the strongest results on Amazon Clothing, Amazon Books, Amazon Electronics, Amazon Sports, and Google Reviews, with especially large gains on Books (0.291 \textit{N@10}, 0.466 \textit{R@10}) and solid improvements on the other out-of-domain benchmarks. This shows that optimizing the profiler with embedding-space reward produces user summaries that are more aligned with retrieval utility than untrained profiles or profiles trained with RLAIF, RLPF, or LangPTune. Second, combining recent interactions with a generated profile yields the strongest performance across all datasets. \textit{10 H+\ourmethod{} Profile} achieves the best results on every benchmark, reaching 0.173/0.306 on Clothing, 0.329/0.516 on Books, 0.143/0.267 on Electronics, 0.198/0.376 on Sports, and 0.264/0.436 on Google Reviews. This indicates that recent interaction history and learned profile summaries provide complementary information: the former captures short-term intent, while the latter summarizes more stable and transferable preferences. Third, the gains of \ourmethod{} are particularly clear on out-of-domain datasets, which suggests that the profiles learned by \ourmethod{} generalize better across domains than raw histories alone. Overall, these results show that \ourmethod{} can improve sequential recommendation quality even when the embedding model is frozen, which is especially useful for proprietary or off-the-shelf embedding services that cannot be further tuned.

\begin{table}[t]
   % \vspace{-0.3cm}
    % \caption{Embedding-Adaptive sequential recommendation results when the embedding model is trained on Amazon Clothing and directly evaluated on other domains without target-domain fine-tuning. \textit{N@10} and \textit{R@10} denote NDCG@10 and Recall@10, respectively, and higher values indicate better performance.}
    \caption{Embedding-adaptive sequential recommendation results. The embedding model is trained on Amazon Clothing with supervised contrastive learning and directly evaluated on other domains. \textit{N@10} and \textit{R@10} denote NDCG@10 and Recall@10. Higher is better.}
    \centering
   \small
   \begin{adjustbox}{max width=1\linewidth}
   \begin{tabular}{lcc|cccccccc}
   \toprule[1.2pt]
   \multirow{2}{*}{\textbf{Method}} & \multicolumn{2}{c}{\cellcolor{blue!20}\textbf{Amazon Clothing}} & \multicolumn{2}{c}{\cellcolor{orange!20}\textbf{Amazon Books}} & \multicolumn{2}{c}{\cellcolor{orange!20}\textbf{Amazon Electronics}} & \multicolumn{2}{c}{\cellcolor{orange!20}\textbf{Amazon Sports}} & \multicolumn{2}{c}{\cellcolor{orange!20}\textbf{Google Reviews}} \\
   \cmidrule(lr){2-3} \cmidrule(lr){4-5} \cmidrule(lr){6-7} \cmidrule(lr){8-9} \cmidrule(lr){10-11}
   & \textit{N@10} & \textit{R@10} & \textit{N@10} & \textit{R@10} & \textit{N@10} & \textit{R@10}& \textit{N@10} & \textit{R@10}& \textit{N@10} & \textit{R@10} \\
   \midrule[0.75pt]
   \textsc{SASRec-Sem} & \underline{0.259} & \textbf{0.526} & 0.159 & 0.371 & 0.128 & 0.287 & 0.180 & 0.384 & 0.167 & 0.340 \\
   \textsc{Bert4Rec-Sem} & \textbf{0.267} & \underline{0.512} & 0.176 & 0.367 & 0.121 & 0.272 & 0.156 & 0.324 & 0.123 & 0.278 \\
   \textsc{UnisRec} & 0.257 & 0.515 & 0.149 & 0.343 & 0.125 & 0.287 & 0.176 & 0.368 & 0.162 & 0.339 \\
   \textsc{10H+SCL} & 0.208 & 0.366 & 0.298 & 0.465 & 0.168 & 0.298 & 0.227  & 0.413 & 0.280 & 0.446 \\
   \textsc{Base+SCL} & 0.210 & 0.365 & 0.311 & \underline{0.483} & \underline{0.181} & \underline{0.312} & \underline{0.235} & {0.418} & 0.278 & 0.448\\
   \textsc{RLAIF+SCL} & 0.185 & 0.317 & \textbf{0.330} & \textbf{0.500} & 0.160 & 0.287 & 0.226 & 0.403 & 0.226 & 0.404 \\
   \textsc{RLPF+SCL} & 0.210 & 0.370 & 0.301 & 0.463 & 0.166 & 0.288 & 0.229 & 0.410 & {0.284} & \underline{0.457} \\
   \textsc{LangPTune+SCL} & 0.215 & 0.364 & 0.304 & 0.473 & 0.174 & 0.301 & 0.236 & \underline{0.422} & \underline{0.293} & 0.460 \\
   \rowcolor{gray!20}\textsc{\ourmethod{}+SCL} (Ours) & 0.240 & 0.406 &  \underline{0.323} & \textbf{0.500} & \textbf{0.186$^*$} & \textbf{0.325$^*$} & \textbf{0.238} & \textbf{0.424} & \textbf{0.308$^*$} & \textbf{0.484$^*$} \\
   \bottomrule[1.2pt]
   \end{tabular}
   \end{adjustbox}
   \vspace{-0.3cm}
   \label{tab:train_emb}
\end{table}

\paragraph{Embedding-Adaptive Sequential Recommendation.}
Table~\ref{tab:train_emb} reports zero-shot cross-domain retrieval results when the embedding model is further trained with supervised contrastive learning on Amazon Clothing and then directly transferred to other domains. We find that \ourmethod{} remains strong in this more competitive setting. Among the profile-based methods, \textit{\ourmethod{}+SCL} achieves the best overall transfer performance, obtaining the strongest results on Amazon Electronics (0.186 \textit{N@10}, 0.325 \textit{R@10}), Amazon Sports (0.238 \textit{N@10}, 0.424 \textit{R@10}), and Google Reviews (0.308 \textit{N@10}, 0.484 \textit{R@10}), while also matching the best \textit{R@10} on Amazon Books. These gains over \textit{10H+SCL}, \textit{Base+SCL}, \textit{RLAIF+SCL}, \textit{RLPF+SCL}, and \textit{LangPTune} suggest that the profiles produced by \ourmethod{} provide a more informative training signal for embedding adaptation than raw history or alternative profile optimization objectives. Although sequential recommendation baselines such as \textit{SASRec-Sem}, \textit{Bert4Rec-Sem}, and \textit{UnisRec} remain competitive on the source-domain Amazon Clothing split, their transfer performance is generally weaker than profile-based embedding methods on the target domains. This pattern suggests that learned textual profiles capture higher-level and more transferable user preference structure, which helps the adapted embedding model generalize beyond the source domain. Together, these results show that \ourmethod{} is effective not only when the embedding model is frozen, but also when the embedding space is further optimized, leading to stronger cross-domain retrieval after adaptation.

\begin{table}[t]
   % \vspace{-0.3cm}
   % \caption{Experiment results of personalized question answering using user profiles and user history as personalized context. \textit{Next Item} denotes next-item multiple-choice prediction accuracy, where higher values are better. \textit{Rating} denotes user rating prediction measured by mean absolute error, where lower values are better. }
   \caption{Personalized question answering results using user profiles and histories as context. \textit{Next Item} denotes next-item multiple-choice accuracy, where higher is better. \textit{Rating} denotes rating prediction MAE, where lower is better.}
   \centering
   \small
   \begin{adjustbox}{max width=1\linewidth}
   \begin{tabular}{lcc|cccccccc}
   \toprule[1.2pt]
   \multirow{2}{*}{\textbf{Method}} & \multicolumn{2}{c}{\cellcolor{blue!20}\textbf{Amazon Clothing}} & \multicolumn{2}{c}{\cellcolor{orange!20}\textbf{Amazon Books}} & \multicolumn{2}{c}{\cellcolor{orange!20}\textbf{Amazon Electronics}} & \multicolumn{2}{c}{\cellcolor{orange!20}\textbf{Amazon Sports}} & \multicolumn{2}{c}{\cellcolor{orange!20}\textbf{Google Reviews}} \\
   \cmidrule(lr){2-3} \cmidrule(lr){4-5} \cmidrule(lr){6-7} \cmidrule(lr){8-9} \cmidrule(lr){10-11}
   & \textit{Next Item}$\uparrow$ & \textit{Rating}$\downarrow$ & \textit{Next Item}$\uparrow$ & \textit{Rating}$\downarrow$ & \textit{Next Item}$\uparrow$ & \textit{Rating}$\downarrow$ & \textit{Next Item}$\uparrow$ & \textit{Rating}$\downarrow$ & \textit{Next Item}$\uparrow$ & \textit{Rating}$\downarrow$ \\
   \midrule[0.75pt]
   \textsc{Base}  & 0.327 & 1.698 & 0.514 & 1.643 & 0.308 & 1.788 & 0.306 & 1.500 & 0.273 & 1.966 \\
   \textsc{RLAIF} & \underline{0.415} & 1.651 & 0.553 & 1.785 & 0.376 & 1.958 & \underline{0.438} & 1.814 & 0.250 & 2.143 \\
   \textsc{RLPF} & 0.377 & \underline{1.599} & 0.544 & 1.607 & 0.384 & \textbf{1.655} & 0.381 & \textbf{1.471} & 0.248 & 1.913 \\
   \textsc{LangPTune} & 0.313 & 1.640 & \underline{0.570} & \textbf{1.493} & \underline{0.418} & 1.667 & 0.286 & 1.492 & \underline{0.286} & \underline{1.885} \\
   \rowcolor{gray!20}\textsc{\ourmethod{}} (Ours) & \textbf{0.432$^*$} & \textbf{1.598} & \textbf{0.582} & \underline{1.535} & \textbf{0.425} & \underline{1.656} & \textbf{0.452$^*$} & \underline{1.491} & \textbf{0.351$^*$} & \textbf{1.823$^*$} \\
   \hline
   \textsc{All History} & 0.438 & \underline{1.047} & 0.592 & \textbf{0.864} & 0.515 & \underline{1.102} & 0.556 & \underline{1.040} & 0.221 & \textbf{0.835} \\
   \textsc{10 History} & 0.430 & 1.263 & 0.595 & 1.026 & 0.423 & 1.238 & 0.460 & 1.210 & 0.283 & \underline{1.075} \\
   \textsc{10 H+Base} & 0.504 & 1.236 & 0.602 & 1.237 & 0.483 & 1.351 & 0.532 & 1.192 & 0.386 & 1.547 \\
   \textsc{10 H+RLAIF} & 0.493 & 1.409 & 0.598 & 1.308 & 0.498 & 1.484 & 0.547 & 1.378 & 0.385 & 1.737 \\
   \textsc{10 H+RLPF} & \underline{0.526} & 1.240 & 0.612 & 1.164 & \underline{0.521} & 1.326 & 0.564 & 1.152 & 0.369 & 1.532 \\
   \textsc{10 H+LangPTune} & 0.442 & 1.170 & \underline{0.636} & 1.151 & 0.496 & 1.283 & \underline{0.567} & 1.132 & \underline{0.454} & 1.402 \\
   \rowcolor{gray!20}\textsc{10 H+\ourmethod{} (Ours)} & \textbf{0.557$^*$} & \textbf{1.020} & \textbf{0.669$^*$} & \underline{0.867} & \textbf{0.545$^*$} & \textbf{1.093} & \textbf{0.592$^*$} & \textbf{0.876$^*$} & \textbf{0.510$^*$} & 1.109 \\
   \bottomrule[1.2pt]
   \end{tabular}
   \end{adjustbox}
   \vspace{-0.3cm}
   \label{tab:pqa}
\end{table}

\paragraph{Personalized Question Answering.}
Table~\ref{tab:pqa} evaluates whether the generated profiles serve as effective natural-language context for downstream personalized question answering. The results show that \ourmethod{} improves both the quality of the profile itself and its usefulness for personalized QA. In the profile-only setting, \textit{\ourmethod{}} achieves the best next-item prediction accuracy on all five datasets and also attains the lowest rating prediction MAE on Amazon Books, Amazon Sports, and Google Reviews, outperforming competing profile generation methods such as \textit{Base}, \textit{RLAIF}, \textit{RLPF}, and \textit{LangPTune}. For example, on Google Reviews, \textit{\ourmethod{}} improves next-item accuracy from 0.250 and 0.248 under \textit{RLAIF} and \textit{RLPF} to 0.351, while also reducing MAE to 1.823. This suggests that the proposed reward improves not only retrieval-oriented utility, but also the faithfulness and informativeness of the generated profile in text space. When combined with recent user behavior, the gains become even larger. \textit{10 History+\ourmethod{}} achieves the best next-item accuracy on all five domains and also delivers the best or near-best rating prediction error, reaching 0.557/1.020 on Clothing, 0.669/0.867 on Books, 0.545/1.093 on Electronics, 0.592/0.876 on Sports, and 0.510/1.109 on Google Reviews. Compared with both raw-history baselines and other hybrid profile methods, these results indicate that a learned profile complements recent interactions by summarizing stable user preferences in a compact and usable form. Overall, the personalized QA results show that \ourmethod{} produces profiles that are not only effective for embedding-based sequential recommendation, but also interpretable and practically useful for downstream LLM-based generative personalization.

\section{Analysis}

\begin{table}[t]
   % \vspace{-0.3cm}
   \caption{In-domain sequential recommendation performance under the trained embedding model setting. Models are trained and evaluated within each domain. The final two columns report the average performance across all five domains.}
   \centering
   \small
   \begin{adjustbox}{max width=1\linewidth}
   \begin{tabular}{lcccccccccc|cc}
   \toprule[1.2pt]
   \multirow{2}{*}{\textbf{Method}} & \multicolumn{2}{c}{\cellcolor{blue!20}\textbf{Amazon Clothing}} & \multicolumn{2}{c}{\cellcolor{blue!20}\textbf{Amazon Books}} & \multicolumn{2}{c}{\cellcolor{blue!20}\textbf{Amazon Electronics}} & \multicolumn{2}{c}{\cellcolor{blue!20}\textbf{Amazon Sports}} & \multicolumn{2}{c}{\cellcolor{blue!20}\textbf{Google Reviews}} & \multicolumn{2}{c}{\cellcolor{magenta!20}\textbf{Average}}\\
   \cmidrule(lr){2-3} \cmidrule(lr){4-5} \cmidrule(lr){6-7} \cmidrule(lr){8-9} \cmidrule(lr){10-11} \cmidrule(lr){12-13} 
   & \textit{N@10} & \textit{R@10} & \textit{N@10} & \textit{R@10} & \textit{N@10} & \textit{R@10}& \textit{N@10} & \textit{R@10} & \textit{N@10} & \textit{R@10} & \textit{N@10} & \textit{R@10} \\
   \midrule[0.75pt]
   \textsc{SASRec} & \textbf{0.364} & \textbf{0.569} & 0.249 & 0.381 & \underline{0.348} & 0.514 & 0.211 & 0.358 & 0.274 & 0.461 & 0.289 & 0.457\\
\textsc{Bert4Rec} & \underline{0.355} & 0.520 & 0.244 & 0.375 & \textbf{0.362} & \underline{0.524} & 0.227 & 0.387 & \textbf{0.454} & \textbf{0.661} & \textbf{0.328} & 0.493\\
\textsc{SaSRec-Sem} & 0.259 & \underline{0.526} & 0.351 & \underline{0.635} & 0.329 & \textbf{0.581}  & \underline{0.298} & 0.563 & 0.302 & 0.565 & 0.308 & \underline{0.574}\\
\textsc{Bert4Rec-Sem} & 0.267 & 0.512 & \underline{0.377} & 0.628 & 0.272 & 0.501 & \textbf{0.329} & \textbf{0.607} & 0.309 & \underline{0.647} & \underline{0.311} & \textbf{0.579}\\
\textsc{UnisRec} & 0.257 & 0.515 & \textbf{0.384} & \textbf{0.667} & 0.243 & 0.498 & 0.284 & \underline{0.569} & \underline{0.314} & 0.572 & 0.296 & 0.564\\
\hline
\textsc{10H+SCL} & 0.208 & 0.366 & 0.388 & 0.588 & \textbf{0.212} & \textbf{0.360} & 0.239 & 0.415 & 0.497 & 0.745 & 0.309 & 0.495\\
\textsc{Base+SCL} & {0.210} & 0.365 & 0.396 & 0.594 & \underline{0.204} & \underline{0.355} & \underline{0.253} & \underline{0.440} & 0.495 & 0.750 & \underline{0.312} & \underline{0.501}\\
\textsc{RLAIF+SCL} & 0.185 & 0.317 & \underline{0.407} & \underline{0.616} & 0.180 & 0.315 & 0.249 & 0.434 & 0.491 & 0.739 & 0.302 & 0.484\\
\textsc{RLPF+SCL} & {0.210} & \underline{0.370} & \underline{0.407} & 0.588 & 0.184 & 0.323 & 0.245 & 0.423 & \underline{0.503} & \underline{0.756} & 0.310 & 0.492\\
\textsc{LangPTune+SCL} & \underline{0.215} & 0.364 & 0.406 & \underline{0.616} & 0.180 & 0.318 & \underline{0.253} & 0.435 & 0.494 & 0.747 & 0.310 & 0.496 \\
\rowcolor{gray!20}\textsc{\ourmethod{}+SCL} (Ours) & \textbf{0.240} & \textbf{0.406} & \textbf{0.423$^*$} & \textbf{0.629} & 0.200 & 0.347 & \textbf{0.262} & \textbf{0.458} & \textbf{0.525$^*$} & \textbf{0.780$^*$} & \textbf{0.330} & \textbf{0.524} \\
   \bottomrule[1.2pt]
   \end{tabular}
   \end{adjustbox}
   \vspace{-0.3cm}
   \label{tab:in_domain}
\end{table}

% \subsection{In-Domain Sequential Recommendation Performance}
% \label{in-domain}
% We further examine whether the learned profiles remain useful when the embedding model is trained and evaluated within the same domain, rather than transferred across domains. Table~\ref{tab:in_domain} shows that \textit{\ourmethod{}+SCL} achieves the best average performance across all five datasets, reaching 0.330 in \textit{N@10} and 0.524 in \textit{R@10}, which suggests that the learned profiles remain effective even after in-domain embedding adaptation. Compared with other profile-based baselines, our method delivers consistent improvements on Amazon Clothing, Amazon Books, Amazon Sports, and Google Reviews, and achieves the strongest overall balance across domains. In particular, it obtains the best results on Amazon Books and Google Reviews, with 0.423/0.629 and 0.525/0.780 in \textit{N@10}/\textit{R@10}, respectively, indicating that the reinforcement-learned profiles provide more useful supervision for embedding training than untrained or weakly supervised alternatives. Although classical sequential recommendation models such as \textit{SASRec} and \textit{Bert4Rec} remain competitive on some individual domains, especially Amazon Electronics, their performance is less consistent across datasets. By contrast, \ourmethod{} achieves the strongest average results, showing that profiles optimized to be useful in both text space and embedding space can support robust recommendation performance not only in transfer settings, but also in standard in-domain training. 

\subsection{In-Domain Sequential Recommendation Performance}
\label{in-domain}
We further examine whether the learned profiles remain useful when the embedding model is trained and evaluated within the same domain, rather than transferred across domains. Table~\ref{tab:in_domain} shows that \textit{\ourmethod{}+SCL} achieves the best average performance across all five datasets, reaching 0.330 in \textit{N@10} and 0.524 in \textit{R@10}, which suggests that the learned profiles remain effective even after in-domain embedding adaptation. Compared with other profile-based baselines, our method delivers consistent improvements on Amazon Clothing, Amazon Books, Amazon Sports, and Google Reviews, and achieves the strongest overall balance across domains. In particular, it obtains the best results on Amazon Books and Google Reviews, with 0.423/0.629 and 0.525/0.780 in \textit{N@10}/\textit{R@10}, respectively, indicating that the reinforcement-learned profiles provide more useful supervision for embedding training than untrained or weakly supervised alternatives. Although classical sequential recommendation models such as \textit{SASRec} and \textit{Bert4Rec} remain competitive on some individual domains, especially Amazon Electronics, their performance is less consistent across datasets. A more detailed analysis of these domain-specific differences is provided in Appendix~\ref{indomain_analysis}, where we study how recent-dominance patterns help explain the variation in in-domain performance. By contrast, \ourmethod{} achieves the strongest average results, showing that profiles optimized to be useful in both text space and embedding space can support robust recommendation performance not only in transfer settings, but also in standard in-domain training.

% \subsection{In-Domain Sequential Recommendation Performance}
% Table~\ref{tab:in_domain} reports the in-domain recommendation results when the embedding model is trained and evaluated within the same domain. We observe that \textsc{\ourmethod{}+SCL} achieves the best average performance across all five datasets, reaching 0.330 in \textit{N@10} and 0.524 in \textit{R@10}, which confirms that the learned user profiles remain highly effective even when the model is optimized in-domain. Compared with other profile-based baselines, our method obtains consistent gains on Amazon Clothing, Amazon Books, Amazon Sports, and Google Reviews, and achieves the strongest overall balance across domains. In particular, it gives the best results on Amazon Books and Google Reviews with 0.423/0.629 and 0.525/0.780 in \textit{N@10}/\textit{R@10}, respectively, showing that the reinforcement-learned profile provides a better user representation than untrained or weakly supervised profiles during embedding model training. Although classical sequential recommendation models such as \textsc{SASRec} and \textsc{Bert4Rec} remain competitive on some domains, especially Amazon Electronics, their advantage is less consistent across datasets. By contrast, our method delivers the strongest average performance, suggesting that optimizing textual user profiles to align with both natural language semantics and embedding-space ranking signals leads to more robust in-domain recommendation performance.

\begin{figure}[t]
    \centering
    \begin{minipage}[t]{0.48\textwidth}
        \centering
        \includegraphics[width=\linewidth]{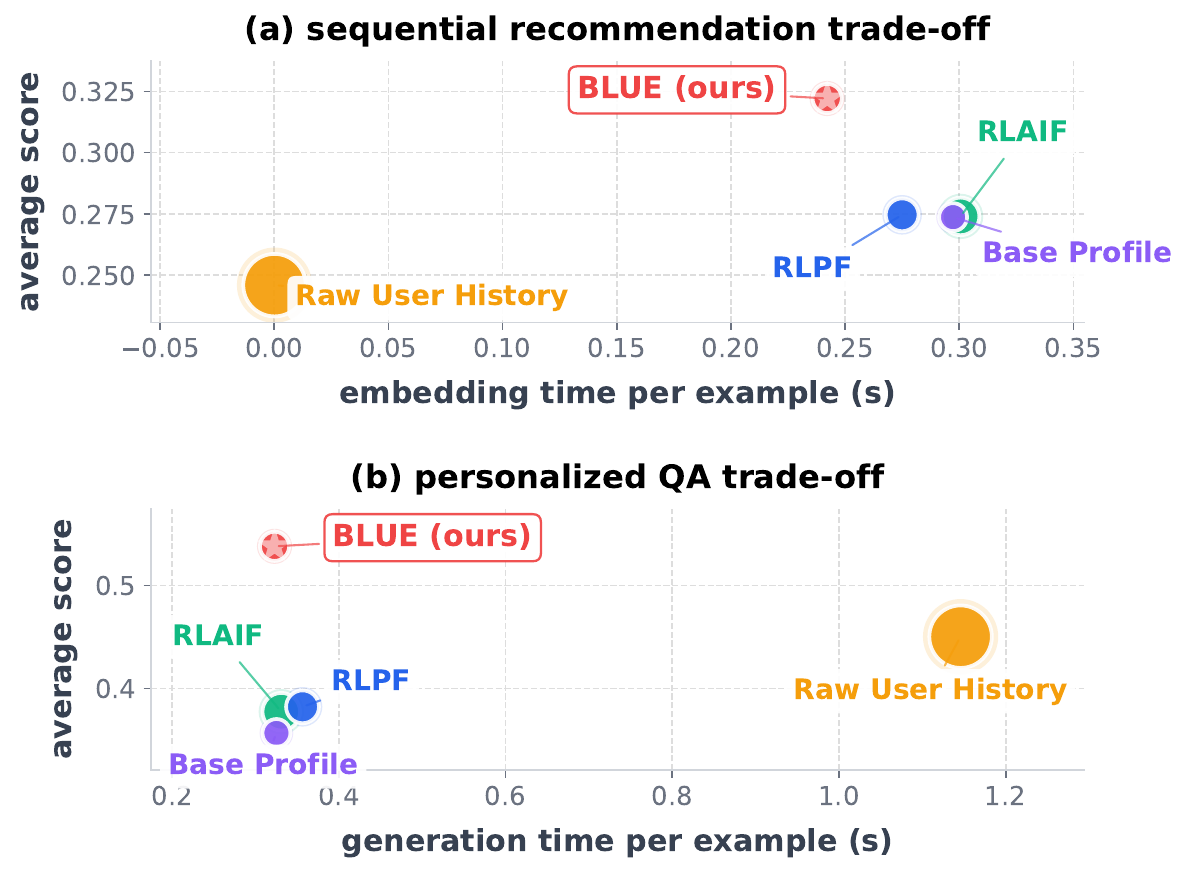}
        \caption{Quality-efficiency trade-off of different user context construction methods. The x-axis shows total inference time per example, and the y-axis shows the overall quality score computed from both sequential recommendation settings and personalized question answering performance. Higher and more leftward points indicate better practical trade-offs.}
        \label{fig:efficiency}
    \end{minipage}
    \hfill
    \begin{minipage}[t]{0.48\textwidth}
        \centering
        \includegraphics[width=0.9\linewidth]{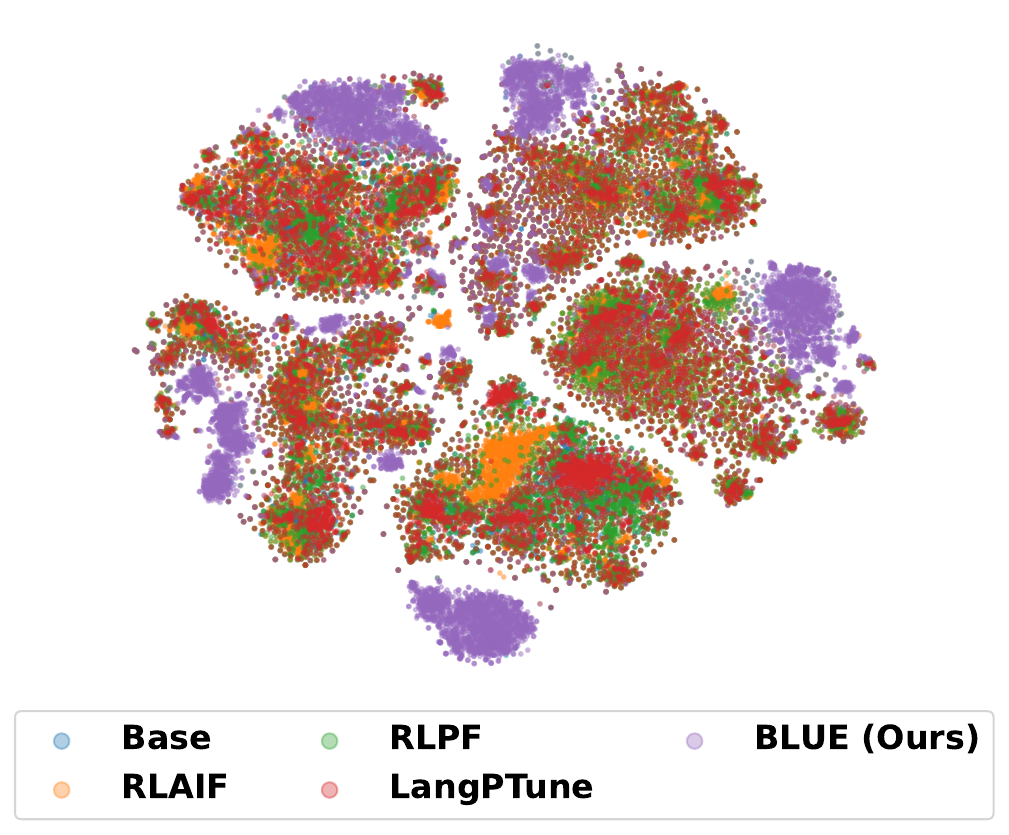}
    \caption{T-SNE visualization of user embeddings induced by profiles generated by different methods under a frozen embedding model. Compared with baseline profile construction methods, \ourmethod{} yields embeddings with tighter local clusters and clearer separation across user groups, suggesting a more discriminative and preference-aware latent structure.}
    \label{fig:tsne}
    \end{minipage}
\end{figure}

% \begin{figure}
%     \centering
%     \includegraphics[width=0.7\linewidth]{assets/efficiency_tradeoff_two_panels.pdf}
%     \caption{Quality-efficiency trade-off of different user context construction methods. The x-axis shows total inference time per example, and the y-axis shows the overall quality score computed from query-adaptive retrieval and personalized question answering performance. Higher and more leftward points indicate better practical trade-offs.}
%     \label{fig:efficiency}
% \end{figure}

\subsection{Efficiency Analysis}
Efficiency analysis is important in our setting because a practical personalization method should improve both recommendation and question answering quality without adding too much inference cost.  We define the quality score in sequential recommendation as $\text{Quality}=(\text{N@10}+\text{R@10})/2$, where \textit{N@10} and \textit{R@10} come from the query-adaptive sequential recommendation setting with a frozen embedding model. For personalized question answering $\text{Quality}=\left(\text{Acc}+\frac{1}{1+\text{MAE}}\right)/2$, , and \textit{Acc} and \textit{MAE} come from accuracy of next item prediction and MAE of the user rating prediction. As shown in Figure \ref{fig:efficiency}, \textit{Raw User History} is the fastest method because it does not require profile generation, but it also gives the lowest quality score, which suggests that raw interaction sequences are efficient yet weak at organizing user preference signals for both retrieval and downstream applications. Profile-based methods require more inference time, but they achieve better quality by compressing and denoising user history into a more informative textual summary. Among them, \textit{Base Profile}, \textit{RLAIF}, and \textit{RLPF} have similar runtime because they all add one profile generation step, yet their quality remains clearly below \ourmethod{}. In contrast, \ourmethod{} reaches the highest quality score while staying in the same general efficiency range as the other profile-based methods, showing that its gains come from better profile optimization rather than a much more expensive inference pipeline. Overall, the results show that \ourmethod{} provides the best balance between effectiveness and efficiency, making it the strongest operating point on the quality-efficiency frontier.

% \begin{figure}
%     \centering
%     \includegraphics[width=0.5\linewidth]{assets/tsne_visualization.pdf}
%     \caption{T-SNE visualization of user embeddings induced by profiles generated with different methods. Compared with baseline profile construction methods, \ourmethod{} produces embeddings with more compact local clusters and clearer separation between user groups, indicating a more discriminative and preference-aware latent structure.}
%     \label{fig:tsne}
% \end{figure}

\subsection{User Embedding Visualization}
To better understand how different profile construction methods shape the latent user representation space, we visualize the induced user embeddings with T-SNE \citep{van2008visualizing}. Figure~\ref{fig:tsne} presents a T-SNE visualization of user embeddings induced by profiles generated from different methods. We observe that the embeddings produced by different profile construction strategies occupy similar global regions, which suggests that all methods preserve the coarse preference structure of users. However, the local geometry differs noticeably. The embeddings induced by the base profile, RLAIF, RLPF, and LangPTune are more diffusely mixed within the major clusters, indicating that these profiles still contain substantial noise and less discriminative preference signals. In contrast, the embeddings derived from \ourmethod{} show more compact local grouping and clearer separation along several cluster boundaries. This pattern suggests that our learned profiles better organize users with similar preferences while pushing apart users with different behavioral tendencies. Such improved local structure is consistent with our quantitative results in both retrieval and personalized question answering: by removing irrelevant details and preserving preference-critical signals, \ourmethod{} yields user representations that are more discriminative in the latent space while remaining semantically meaningful in text space.

\begin{figure}
    \centering
    \includegraphics[width=0.9\linewidth]{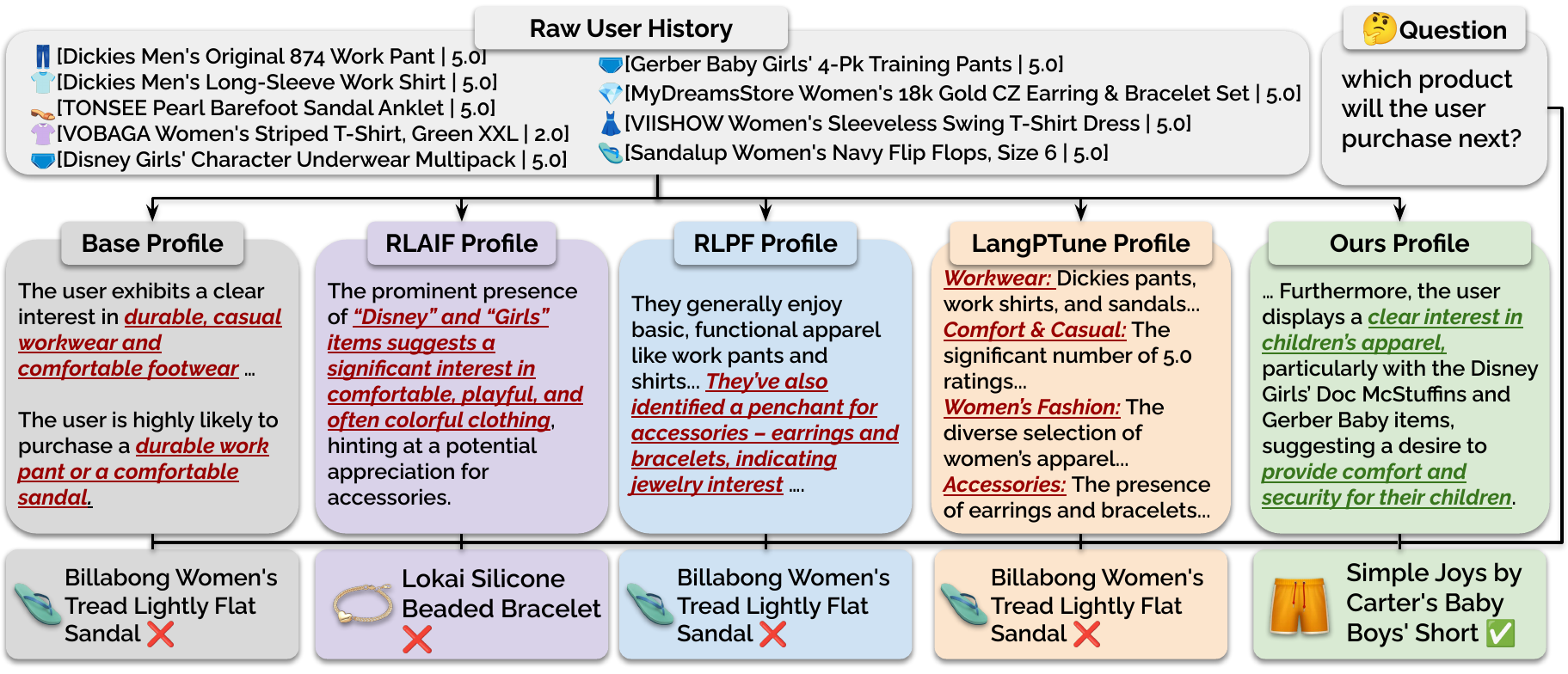}
    \caption{Case study comparing profiles generated by different methods. Baselines mainly focus on workwear, sandals, or accessories and predict the wrong next item, while \ourmethod{} captures the key children's apparel signal and correctly predicts the baby clothing product.}
    \label{fig:casestudy}
\end{figure}

\subsection{Case Study}
We present a qualitative case study to illustrate how \ourmethod{} improves user profile quality compared to baseline methods. Figure~\ref{fig:casestudy} shows an example user from the Amazon Clothing domain, where the raw interaction history contains a mixture of workwear items (e.g., pants and long-sleeve shirts), accessories (e.g., jewelry and anklets), and casual clothing (e.g., dresses and T-shirts), making it difficult for raw history or weakly optimized profiles to identify the dominant preference. The base profile produces a generic and surface-level summary, often listing item categories without distinguishing consistent preferences from occasional behaviors, while RLAIF and RLPF improve fluency and coherence but still introduce noise by over-emphasizing infrequent categories, such as children’s clothing, or by failing to separate primary style from auxiliary purchases. In contrast, \ourmethod{} generates a more structured and selective profile that captures the user’s core preference for practical and casual workwear while treating accessories as secondary interests, effectively filtering out noisy signals and highlighting consistent behavioral patterns, which leads to better alignment with the ground-truth next item. As a result, the induced embedding is closer to relevant items in the latent space, improving retrieval performance, while the profile remains concise and interpretable, providing clearer context for downstream personalized question answering. This example shows that joint optimization in text space and embedding space helps distinguish stable preferences from incidental behaviors, producing user profiles that are both more accurate for recommendation and more useful for downstream applications.

\begin{table*}[t]
   \caption{Ablation study of \ourmethod{} on both sequential recommendation and personalized question answering with both recent 10 history and generated user profile. For retrieval, we report results under the frozen embedding model setting.}
   \centering
   \small
   \begin{adjustbox}{max width=0.9\textwidth}
   \begin{tabular}{lcccccc|cccccc}
   \toprule[1.2pt]
   \multirow{2}{*}{\textbf{Method}} 
   & \multicolumn{6}{c}{\textbf{Sequential Recommendation}} 
   & \multicolumn{6}{c}{\textbf{Personalized Question Answering}} \\
   \cmidrule(lr){2-7} \cmidrule(lr){8-13}
   & \multicolumn{2}{c}{\cellcolor{blue!20}\textbf{Clothing}}
   & \multicolumn{2}{c}{\cellcolor{orange!20}\textbf{Books}}
   & \multicolumn{2}{c}{\cellcolor{orange!20}\textbf{Google Reviews}}
   & \multicolumn{2}{c}{\cellcolor{blue!20}\textbf{Clothing}}
   & \multicolumn{2}{c}{\cellcolor{orange!20}\textbf{Books}}
   & \multicolumn{2}{c}{\cellcolor{orange!20}\textbf{Google Reviews}} \\
   \cmidrule(lr){2-3} \cmidrule(lr){4-5} \cmidrule(lr){6-7}
   \cmidrule(lr){8-9} \cmidrule(lr){10-11} \cmidrule(lr){12-13}
   & \textit{N@10} & \textit{R@10}
   & \textit{N@10} & \textit{R@10}
   & \textit{N@10} & \textit{R@10}
   & \textit{Next Item}$\uparrow$ & \textit{Rating}$\downarrow$
   & \textit{Next Item}$\uparrow$ & \textit{Rating}$\downarrow$
   & \textit{Next Item}$\uparrow$ & \textit{Rating}$\downarrow$ \\
   \midrule[0.75pt]
    \rowcolor{gray!20}\textsc{Full \ourmethod{}} & \underline{0.173} & \textbf{0.306} & \textbf{0.329} & \textbf{0.516} & \underline{0.264} & \underline{0.436} & \textbf{0.557} & \textbf{1.020} & \textbf{0.669} & \textbf{0.876} & \textbf{0.510} & \textbf{1.109} \\
   \textsc{w/o $R_{\mathrm{emb}}$}        & 0.136 & 0.252 & 0.252 & 0.412 & 0.230 & 0.393 & 0.501 & 1.211  & 0.619 & 1.243 & 0.405 & 1.540 \\
   \textsc{w/o $R_{\mathrm{text}}$} & {0.172} & \underline{0.300} & \textbf{0.329} & \textbf{0.516} & \textbf{0.265} & \textbf{0.445}  & 0.510 & {1.128} & {0.633} & 1.111 & 0.484 & 1.436  \\
   \textsc{w/o $e_u^{(1)}$} & 0.171 & 0.296 & {0.323} & {0.499} & 0.253 & 0.424 & 0.534 &  1.193 & 0.635 & 1.151 & 0.466 & 1.505 \\ %alpha1
   \textsc{w/o $e_u^{(2)}$}       & 0.161 & 0.295 & 0.292 & 0.477 & 0.253 & 0.430 & 0.529 & 1.134 & 0.628 & {1.111} & 0.421 & {1.433} \\ %alpha0
   \textsc{w/o $R_{U \to I}$} & 0.134 & 0.252 & 0.252 & 0.411 & 0.230 & 0.393 &  0.515 & \underline{1.063} & 0.628 & \underline{1.103} & 0.441 & \underline{1.377} \\
   \textsc{w/o $R_{I \to U}$} & 0.162 & 0.286 & 0.285 & 0.464 & 0.241 & 0.401 & 0.524 & 1.224 & \underline{0.634} & 1.216 & 0.451 & 1.473 \\
   % \textsc{w/o KL} & \\
   \textsc{w/ REINFORCE++} & \textbf{0.176} & \underline{0.300} & \underline{0.326} & \underline{0.511} & 0.249 & 0.411 & \underline{0.542} & 1.193 & 0.632 & 1.211 & \underline{0.493} & 1.496 \\
   \bottomrule[1.2pt]
   \end{tabular}
   \end{adjustbox}
   \vspace{-0.3cm}
   \label{tab:ablation}
\end{table*}

\subsection{Ablation Study}
We conduct an ablation study to examine how each component of \ourmethod{} affects sequential recommendation and personalized question answering. As shown in Table~\ref{tab:ablation}, the two reward terms play distinct but complementary roles. Removing the embedding-space reward $R_{\mathrm{emb}}$ causes the largest and most consistent degradation, with sharp drops in frozen-embedding retrieval across Clothing, Books, and Google Reviews, together with substantial declines in personalized QA, especially for rating prediction, suggesting that latent-space supervision is the main signal for learning profiles that induce strong user representations. In contrast, removing the text-space reward $R_{\mathrm{text}}$ has only a limited effect on retrieval, and even slightly improves retrieval on Google Reviews, but it consistently weakens personalized QA across all domains, indicating that this reward mainly helps preserve profile content that is useful for downstream language understanding and reasoning. Removing either the $e_u^{(1)}$ reward or the $e_u^{(2)}$ reward also reduces performance, showing that the two embedding views provide complementary supervision: $e_u^{(1)}$ helps retain short-term behavioral evidence by combining recent history with the generated profile, while $e_u^{(2)}$ encourages the profile itself to encode sufficient preference information as a compact standalone summary. We also find that removing $R_{U \to I}$ leads to a larger retrieval drop than removing $R_{I \to U}$, while both directions hurt personalized QA, suggesting that the two alignment directions are complementary rather than redundant. Finally, replacing GRPO with REINFORCE++~\citep{hu2025reinforce++} yields weaker results on most metrics, with especially clear drops on personalized QA, although it remains competitive on a few retrieval numbers. Overall, the full model achieves the best balance between retrieval quality and personalized reasoning, which supports the value of jointly optimizing user profiles in both embedding space and text space.

\section{Related Work}

\paragraph{Latent User Representation Learning for Recommendation.}
Sequential recommendation has long relied on latent user representations learned from historical interaction sequences~\citep{harte2023leveraging,boz2025improving}. Early neural approaches focused on modeling user dynamics with architectures such as recurrent networks \citep{hidasi2015session,yue2024linear,liu2024mamba4rec} and self-attention~\citep{kang2018self,sun2019bert4rec,zhai2024actions}, capturing short- and long-range behavioral dependencies. More recent work has moved beyond pure ID-based modeling and introduced semantic signals from item text or pretrained representations to improve generalization~\citep{liu2025llmemb,he2025llm2rec,hou2024bridging,tan2024personalized,doddapaneni2024user, liu2025uqabench}. For example, UniSRec~\citep{hou2022towards} learns transferable sequence representations from item descriptions for cross-domain recommendation, while LLM-enhanced representation learning methods such as RLMRec~\citep{ren2024representation} align collaborative signals with semantic information derived from language models. Although these methods improve recommendation quality and transferability, they still primarily represent users as dense latent vectors, which remain difficult to inspect or edit directly. Our work is related to this line of research, but differs in that we explicitly optimize a textual user profile and use the embedding model as a reward provider, thereby connecting interpretable language-space summaries with latent retrieval-oriented representations.

\paragraph{Language-Based User Profiles for Personalization.}
A growing line of work studies user representations in natural language rather than only in latent space. Recent methods show that LLMs can summarize a user's historical interactions into compact textual profiles that are easier to inspect, edit, and use for downstream reasoning~\citep{radlinski2022natural,zhou2024language,ramos2024transparent,tan2025can, wang2024llms}. Such profiles have been explored for recommendation, transparency, and scrutability, showing that language-based user modeling can make personalized predictions more understandable and controllable~\citep{ramos2024transparent}. However, most early approaches rely on zero-shot prompting or manually designed profile templates, which makes profile quality highly dependent on prompt design and often weakly tied to downstream utility. To address this, newer methods begin to optimize generated profiles more directly. LangPTune~\citep{langptune} trains LLM-generated user profiles end-to-end for recommendation, and recent profile optimization work studies feedback-driven profile generation under adaptive or task-aware settings~\citep{wu2025rlpf,wang2025lettingo}. Instead of optimizing profiles only from language-space feedback or recommendation labels, our proposed \ourmethod{} jointly optimize them using both text-space supervision and latent embedding-space reward, so that the resulting profile is both interpretable and directly useful for retrieval.

\paragraph{Reinforcement Learning for Language Models.}
Our method is also related to reinforcement learning for optimizing language models under weak or indirect supervision. When there is no single gold output, policy optimization offers a natural way to improve generation using task rewards rather than imitation targets alone. Early reinforcement learning from human feedback (RLHF) work for language models commonly used PPO to optimize responses against learned reward models while constraining policy drift from a reference model~\citep{schulman2017proximal,ouyang2022training}. More recent work has studied simpler or more efficient variants, including GRPO for group-relative reward optimization over multiple sampled outputs~\citep{shao2024deepseekmath}, and critic-free methods such as REINFORCE++~\citep{hu2025reinforce++} and DAPO~\citep{yu2025dapo}. In recommendation and personalization, related studies use reinforcement learning to optimize prompts, summaries, or profile formats according to downstream utility rather than human-written profile references \citep{wu2025rlpf,wang2025lettingo}. Our work extends this line by treating textual user profiling itself as a policy optimization problem, with reward defined jointly in text space through next-item prediction and in latent space through contrastive embedding alignment.

% \paragraph{Reinforcement Learning for Language Models.}
% Our method is also related to reinforcement learning for optimizing language models with indirect supervision. In settings where there is no single gold output, reinforcement learning has become a natural tool for optimizing generation against task-specific rewards rather than imitation targets alone. Group-based policy optimization methods such as GRPO~\citep{shao2024deepseekmath} provide an efficient way to improve language models using relative reward signals across multiple sampled outputs. More recent work has also explored critic-free alternatives such as \textsc{REINFORCE++} for stable language model alignment~\citep{hu2025reinforce++}. In recommendation and personalization, reinforcement learning has been used to optimize prompts, user summaries, or profile formats based on downstream prediction quality instead of human-written profile references \citep{wu2025rlpf,wang2025lettingo}. Our work extends this line by treating textual user profiling itself as a policy optimization problem, where the reward is defined not only by text-space next-item prediction but also by contrastive alignment in the latent embedding space. This design is particularly important for personalization, because there is usually no unique correct natural-language summary for a user, yet there are clear downstream signals indicating whether a profile captures behaviorally useful preference information.

\section{Conclusion}
In this work, we introduced \ourmethod{}, a reinforcement learning framework that bridges textual user profiles and latent user embeddings for personalization. Our method uses a profiler LLM to generate interpretable natural language user profiles, while a latent embedding model provides contrastive reward signals that guide profile generation toward downstream retrieval utility. By combining embedding-space reward with a text-space next-item prediction reward, \ourmethod{} learns profiles that are both semantically meaningful and effective for recommendation. Experiments on Amazon Reviews and Google Local Reviews showed that our method consistently improves query-adaptive retrieval under frozen embedding models, strengthens cross-domain and in-domain recommendation when the embedding model is trained, and provides better personalized context for downstream question answering. These results suggest that textual profiles and latent embeddings should not be viewed as competing forms of user representation, but as complementary spaces that can be jointly optimized to support both interpretability and predictive performance in personalized systems.
% \section*{Acknowledgements}

\bibliography{main}

\appendix

% \newpage

\section{Baseline Details}

\begin{tcolorbox}
{\fontfamily{cmss}\selectfont
A good summary is a shorter piece of text that has the essence of the original. It tries to accomplish the same purpose and conveys the key information from the original input text. Below we define four evaluation axes for summary quality: coherence, accuracy, coverage, and overall quality.\\
~\\
Coherence: This axis answers the question “how coherent is the summary on its own?” A summary is coherent if it's easy to understand when read on its own and free of English errors. A summary is not coherent if it's difficult to understand what the summary is trying to say. Generally, it's more important that the summary is understandable than it being free of grammar errors.

Accuracy: This axis answers the question “does the factual information in the summary accurately match the input user history?” A summary is accurate if it doesn't say things that aren't in the input user hisory and generally is not misleading.

Coverage: This axis answers the question “how well does the summary cover the important information in the input user history?” A summary has good coverage if it mentions the main information from the input user history text that's important to understand the situation described in the user history. A summary has poor coverage if someone reading only the summary would be missing several important pieces of information about the situation in the input text.

Overall quality: This axis answers the question “how good is the summary overall at representing the input user history?” This can encompass all of the above axes of quality, as well as others you feel are important. If it's hard to find ways to make the summary better, the overall quality is good. If there are lots of different ways the summary can be made better, the overall quality is bad.

~\\
You are an expert user summary rater. Given a USER\_HISTORY\_TEXT as user past activity history in text and a SUMMARY that summarizes users' preference based on the past activity history, your role is to provide a SCORE from 1 to 10 that rates the quality of the SUMMARY given the USER\_HISTORY\_TEXT, with 1 being awful and 10 being a perfect SUMMARY.\\
~\\
USER\_HISTORY\_TEXT: \{user\_history\} \\
SUMMARY: \{user\_profile\} \\
SCORE:
}
\end{tcolorbox}
\noindent\begin{minipage}{1.0\textwidth}
\captionof{table}{Prompt used in RLAIF baseline to provide reward signals for generated user profile.}\label{rlaif_prompt}
\end{minipage}

\begin{tcolorbox}
{\fontfamily{cmss}\selectfont
 You are a recommender system.
User profile:
\{user\_profile\}

Question:  Based on this information, which product will the user purchase next? Choose one from the following options.\\
Options:\\
A. \{Product\_Title\_1\}\\
B. \{Product\_Title\_2\}\\
C. \{Product\_Title\_3\}\\
D. \{Product\_Title\_4\}\\
Structure the answer with '[Index of your choice (A/B/C/D)]. Only output the final answer index, no other explanations.
}
\end{tcolorbox}
\noindent\begin{minipage}{1.0\textwidth}
\captionof{table}{Prompt used in RLPF baseline to generate next item prediction using generated user profile.}\label{rlpf_prompt}
\end{minipage}

\begin{itemize}[leftmargin=*]
    \item \textbf{RLAIF} \citep{rlaif}: following the original implementation, after getting the user profile $p_u$ from raw user history $\mathcal{H}_u$, we then feed the user profile into an external LLM for user profile quality scoring, we use \textit{Gemma3-4B-it} in this case. The exact prompt template is shown in Table \ref{rlaif_prompt}. Given the feedback score, we perform GRPO training on the profiler model, while filter out examples that all rollout share the same reward score.  
    \item \textbf{RLPF} \citep{wu2025rlpf}: after getting the user profile $p_u$ from raw user history $\mathcal{H}_u$, we then feed the user profile into an external LLM for user profile quality scoring, we use \textit{Gemma3-4B-it} to perform next item prediction in this case. The next item prediction prompt is shown in Table \ref{rlpf_prompt}, we give four options to the model, where one of them is the ground-truth next interacted item, while other options are randomly sampled from negative items and shuffle the order when feeding into LLMs for prediction. 
    \item \textbf{LangPTune} \citep{langptune}: trThe LangPTune baseline method uses the same Qwen3-emb-0.6B as the embedding model for fair comparison. The main difference is the reward function, where is the NDCG@10 metric by feeding the generated user profile and ranking positive items against negative items. 
\end{itemize}

\section{Task Details}

\begin{tcolorbox}
{\fontfamily{cmss}\selectfont
Given a user profile, retrieve items that the user is most likely to purchase next.\\
\{user\_profile\}
}
\end{tcolorbox}
\noindent\begin{minipage}{1.0\textwidth}
\captionof{table}{Prompt template to encode user profile in sequential recommendation.}\label{profileonly_prompt}
\end{minipage}

\begin{tcolorbox}
{\fontfamily{cmss}\selectfont
Given a user profile, retrieve items that the user is most likely to purchase next.\\
\{user\_history\}\\
\{user\_profile\}
}
\end{tcolorbox}
\noindent\begin{minipage}{1.0\textwidth}
\captionof{table}{Prompt template to encode both user history and user profile in sequential recommendation.}\label{historyprofile_prompt}
\end{minipage}

\begin{tcolorbox}
{\fontfamily{cmss}\selectfont
\textit{\% Amazon Item Template}\\
Product Item Title: \{title\}\\
Product Item Description: \{description\}\\
Product Item Features: \{features\}

~\\
\textit{\% Google Local Review Location Template}\\
Location Name: \{name\}\\
Location Address: \{address\}\\
Location Description: \{description\}\\
Location Category: \{category\}
}

\end{tcolorbox}
\noindent\begin{minipage}{1.0\textwidth}
\captionof{table}{Prompt template to encode item information in sequential recommendation.}\label{item_prompt}
\end{minipage}

\begin{itemize}[leftmargin=*]
\item \textbf{Sequential Recommendation} Given a user context, which is encoded in the template in Table \ref{profileonly_prompt} and \ref{historyprofile_prompt} for profile-only and history+profile sequential recommendation respectively, input into embedding model to obtain user embedding. Item are encoded in the template in Table \ref{item_prompt} and feed into embedding model to obtain item embeddings. The user embedding is then used to calculate cosine similarity with the item embeddings to obtain ranking scores, where higher score rank first. 

\item \textbf{Personalized Question Answering} Given a user context, we ask the model (Gemma3-12B-it) model to predict user next interacted item and predict next item rating, the prompt template is shown in Table~\ref{qa_nextitem_prompt} and \ref{qa_rating_prompt}. The data for personalized question answering is the same as sequential recommendation, while performed in natural language form to test the text-space profile utility.

\end{itemize}

\begin{tcolorbox}
{\fontfamily{cmss}\selectfont
\% \textit{Only User Profile as Context}\\
Here is a summary of the user's preferences (between <summary> and </summary>): <summary>\{user\_profile\}</summary>\\

\% \textit{Recent History + User Profile as Context}\\
Here is the user's recent interaction history (between <history> and </history>): <history>\{user\_history\}</history>\\
Here is a summary of the user's preferences (between <summary> and </summary>): <summary>\{user\_profile\}</summary>\\
        
Based on this information, which product will the user purchase next? Choose one from the following options: \\
A. \{Product\_Title\_1\}\\
B. \{Product\_Title\_2\}\\
C. \{Product\_Title\_3\}\\
D. \{Product\_Title\_4\}\\
Structure the answer with '[Index of your choice (A/B/C/D)]'.\\
Answer:
}
\end{tcolorbox}
\noindent\begin{minipage}{1.0\textwidth}
\captionof{table}{Prompt template to do personalized next item prediction task.}\label{qa_nextitem_prompt}
\end{minipage}

\begin{tcolorbox}
{\fontfamily{cmss}\selectfont
\% \textit{Only User Profile as Context}\\
Here is a summary of the user's preferences (between <summary> and </summary>): <summary>\{user\_profile\}</summary>\\

\% \textit{Recent History + User Profile as Context}\\
Here is the user's recent interaction history (between <history> and </history>): <history>\{user\_history\}</history>\\
Here is a summary of the user's preferences (between <summary> and </summary>): <summary>\{user\_profile\}</summary>\\

Based on this information, what is the user\'s rating on the following product? \{product\_info\}\\
Answer a score between 1 to 5 inclusive, using full-point(1.0) increments. Structure the answer with "[Your choice from (1.0, 2.0, 3.0, 4.0, 5.0)]"
}
\end{tcolorbox}
\noindent\begin{minipage}{1.0\textwidth}
\captionof{table}{Prompt template to do personalized item rating prediction task.}\label{qa_rating_prompt}
\end{minipage}

\section{\ourmethod{} Performance with Different Base Model}

To examine whether the effectiveness of \ourmethod{} depends on the choice of profiler backbone, we repeat the experiments using \textit{Qwen3-1.7B} as the base profiler model instead of \textit{Gemma3-1B-it}. The embedding model, training recipe, reward design, and evaluation protocol remain unchanged. We report the results for embedding-frozen sequential recommendation in Table~\ref{tab:frozen_emb_qwen} and personalized question answering in Table~\ref{tab:pqa_qwen}. These experiments test whether the advantage of \ourmethod{} is tied to a specific base model or reflects a more general gain from our training objective.

\paragraph{Embedding-Frozen Sequential Recommendation with Qwen3-1.7B.}
Table~\ref{tab:frozen_emb_qwen} shows that \ourmethod{} remains consistently strong when built on top of \textit{Qwen3-1.7B}. Under the history+profile setting, \textit{10 H+\ourmethod{} Profile} achieves the best performance on all five benchmarks, with clear gains over competing methods. In particular, it improves from 0.145/0.266 to 0.168/0.301 on Amazon Clothing and from 0.289/0.460 to 0.312/0.493 on Amazon Books compared to \textit{10 H+Base Profile}. Similar improvements are observed on Amazon Electronics, Amazon Sports, and Google Reviews. Compared with \textit{10 H+RLAIF Profile}, \textit{10 H+RLPF Profile}, and \textit{10H+LangPTune}, our method consistently achieves the strongest or near-strongest results across both in-domain and out-of-domain settings. These results suggest that the gains mainly come from the joint optimization in text space and embedding space, rather than from a specific profiler backbone.

\paragraph{Personalized Question Answering with Qwen3-1.7B.}
Table~\ref{tab:pqa_qwen} further shows that the profiles learned by \ourmethod{} remain effective for downstream language-based personalization. In the hybrid setting with recent history, \textit{10 History+\ourmethod{}} achieves the best next-item prediction accuracy on four out of five datasets, including Amazon Clothing (0.529), Amazon Books (0.657), Amazon Electronics (0.548), and Google Reviews (0.534), while remaining competitive on Amazon Sports. For rating prediction, although \ourmethod{} does not always achieve the lowest MAE, it consistently stays close to the best-performing methods and even obtains the best score on Amazon Sports (0.798). Compared with profile-based baselines such as \textit{Base}, \textit{RLAIF}, \textit{RLPF}, and \textit{LangPTune}, our method provides a better balance between next-item accuracy and rating prediction across domains. These findings are consistent with the results using \textit{Gemma3-1B-it}, showing that \ourmethod{} improves the usefulness of textual user profiles in a way that generalizes across different base LLMs.

\begin{table}[t]
   % \vspace{-0.3cm}
   \caption{Experimental results under the embedding-frozen sequential recommendation using \textbf{Qwen3-1.7B} as base model. The embedding model is kept fixed, while different user context construction methods are used for ranking. \textit{N@10} and \textit{R@10} denote NDCG@10 and Recall@10, respectively, and higher values indicate better performance. \colorbox{indomain}{Blue} denotes the source-domain Amazon Clothing benchmark, while \colorbox{ood}{orange} denotes out-of-domain test benchmarks without further notice. }
   \centering
   \small
   \begin{adjustbox}{max width=1\linewidth}
   \begin{tabular}{lcc|cccccccc}
   \toprule[1.2pt]
   \multirow{2}{*}{\textbf{Method}} & \multicolumn{2}{c}{\cellcolor{blue!20}\textbf{Amazon Clothing}} & \multicolumn{2}{c}{\cellcolor{orange!20}\textbf{Amazon Books}} & \multicolumn{2}{c}{\cellcolor{orange!20}\textbf{Amazon Electronics}} & \multicolumn{2}{c}{\cellcolor{orange!20}\textbf{Amazon Sports}} & \multicolumn{2}{c}{\cellcolor{orange!20}\textbf{Google Reviews}} \\
   \cmidrule(lr){2-3} \cmidrule(lr){4-5} \cmidrule(lr){6-7} \cmidrule(lr){8-9} \cmidrule(lr){10-11}
   & \textit{N@10} & \textit{R@10} & \textit{N@10} & \textit{R@10} & \textit{N@10} & \textit{R@10}& \textit{N@10} & \textit{R@10}& \textit{N@10} & \textit{R@10} \\
   \midrule[0.75pt]
   % \multicolumn{11}{c}{ \cellcolor{gray!20}{\textit{Qwen3-Embedding-0.6B}}}\\
   \textsc{All History} & 0.091 & 0.175 & 0.134 & 0.256 & 0.105 & 0.209 & 0.131 & 0.254 & 0.123 & 0.242 \\
   \textsc{10 History} & 0.085 & 0.168 & 0.109 & 0.215 & 0.086 & 0.177 & 0.125 & 0.246 & 0.197 & 0.342 \\
   \textsc{10 H+Base Profile} & 0.145 & 0.266 & \underline{0.289} & \underline{0.460} & 0.120 & 0.235 & \underline{0.192} & \underline{0.363} & 0.225 & 0.382 \\
   \textsc{10 H+RLAIF Profile} & 0.114 & 0.214 & 0.238 & 0.406 & 0.108 & 0.223 & 0.175 & 0.339 & 0.218 & 0.364 \\
   \textsc{10 H+RLPF Profile} & {0.139} & {0.254} & {0.285} & {0.452} & \underline{0.129} & \textbf{0.258} & 0.189 & 0.359 & {0.238} & {0.391} \\
   \textsc{10H+LangPTune} & \underline{0.153} & \underline{0.280} & 0.286 & 0.458 & {0.127} & {0.248} & \underline{0.192} & 0.361 & \underline{0.241} & \underline{0.400}  \\
   \rowcolor{gray!20}\textsc{10 H+\ourmethod{} Profile} (Ours) & \textbf{0.168} & \textbf{0.301} & \textbf{0.312} & \textbf{0.493} & \textbf{0.133} & \underline{0.257} & \textbf{0.203} &  \textbf{0.383} & \textbf{0.255} & \textbf{0.422}\\
  
   \bottomrule[1.2pt]
   \end{tabular}
   \end{adjustbox}
   \vspace{-0.3cm}
   \label{tab:frozen_emb_qwen}
\end{table}

\begin{table}[t]
   % \vspace{-0.3cm}
   \caption{Experiment results of personalized question answering using user profiles and user history as personalized context. \textbf{Qwen3-1.7B} is used as profiler base model, and Gemma3-12B-it is used for question answering. \textit{Next Item} denotes next-item multiple-choice prediction accuracy, where higher values are better. \textit{Rating} denotes user rating prediction measured by mean absolute error, where lower values are better. }
   \centering
   \small
   \begin{adjustbox}{max width=1\linewidth}
   \begin{tabular}{lcc|cccccccc}
   \toprule[1.2pt]
   \multirow{2}{*}{\textbf{Method}} & \multicolumn{2}{c}{\cellcolor{blue!20}\textbf{Amazon Clothing}} & \multicolumn{2}{c}{\cellcolor{orange!20}\textbf{Amazon Books}} & \multicolumn{2}{c}{\cellcolor{orange!20}\textbf{Amazon Electronics}} & \multicolumn{2}{c}{\cellcolor{orange!20}\textbf{Amazon Sports}} & \multicolumn{2}{c}{\cellcolor{orange!20}\textbf{Google Reviews}} \\
   \cmidrule(lr){2-3} \cmidrule(lr){4-5} \cmidrule(lr){6-7} \cmidrule(lr){8-9} \cmidrule(lr){10-11}
   & \textit{Next Item}$\uparrow$ & \textit{Rating}$\downarrow$ & \textit{Next Item}$\uparrow$ & \textit{Rating}$\downarrow$ & \textit{Next Item}$\uparrow$ & \textit{Rating}$\downarrow$ & \textit{Next Item}$\uparrow$ & \textit{Rating}$\downarrow$ &\textit{Next Item}$\uparrow$ & \textit{Rating}$\downarrow$ \\
   \midrule[0.75pt]
    \textsc{All History} & 0.438 & \textbf{1.047} & 0.592 & \textbf{0.864} & 0.515 & \textbf{1.102} & 0.556 & {1.040} & 0.221 &  \textbf{0.835} \\
    \textsc{10 History} & 0.430 & 1.263 & 0.595 & 1.026 & 0.423 & 1.238 &  0.460 & 1.210 & 0.283 & 1.275 \\
   \textsc{10 History+Base} & \underline{0.524} & 1.100 & \underline{0.648} & 1.081 & 0.522 & 1.239 & \underline{0.587} & 0.999 & {0.508} & 1.127 \\
   \textsc{10 History+RLAIF} & 0.461 & 1.114 & 0.640 & {1.017} &  \underline{0.529} & 1.142 & 0.579 & \underline{0.946} & 0.493 & 1.195  \\
   \textsc{10 History+RLPF} &  0.521 & \underline{1.063} & 0.652 & \underline{1.013} & 0.528 & 1.143 & 0.578 & {0.968} & \underline{0.514} & 1.102  \\
   \textsc{10 History+LangPTune} & 0.515 & 1.160 &  0.646 & 1.136 & 0.508 & 1.186 & \textbf{0.591} & 1.073 & 0.503 & 1.186  \\
   \rowcolor{gray!20}\textsc{10 History+\ourmethod{} (Ours)} & \textbf{0.529} & 1.160 & \textbf{0.657} & 1.019 & \textbf{0.548} & \underline{1.164} & \underline{0.587} & \textbf{0.798} & \textbf{0.534} & \underline{1.042}  \\
   \bottomrule[1.2pt]
   \end{tabular}
   \end{adjustbox}
   \vspace{-0.3cm}
   \label{tab:pqa_qwen}
\end{table}

\begin{table}[t]
   % \vspace{-0.3cm}
   \caption{Recent-dominance score across multiple recommendation domains, where higher score means the target recommendation is more dominanted by the recent history. }
   \centering
   \small
   \begin{adjustbox}{max width=1\linewidth}
   \begin{tabular}{lccccc}
   \toprule[1.2pt]
    \textbf{Domain} & \textbf{Amazon Clothing} & \textbf{Amazon Books} & A\textbf{mazon Electronics} & \textbf{Amazon Sports} & \textbf{Google Reviews} \\
    \midrule[0.75pt]
   \textsc{Recent-Dominance Score} & 0.0314 & 0.0274 & 0.0833 & 0.0344 & 0.00211\\
   \bottomrule[1.2pt]
   \end{tabular}
   \end{adjustbox}
   \vspace{-0.3cm}
   \label{tab:pqa_qwen}
\end{table}

% \section{In-Domain Sequential Recommendation Performance Analysis}

% To study the domain difference in in-domain sequential recommendation performance in Section \ref{in-domain}, we define a recent-dominance score that use the \textit{Qwen3-Emb-0.6B} to embed the target item title and most recent 10 user behavior title and calculate their mean cosine similarity score denoted as $s_{recent}$ and use all previous user behaviors title mean similarity as $s_{long}$, then we calculate the recent-dominance score by $s_{recent} - s_{long}$. 
% Domain-wise differences in in-domain sequential recommendation appear to be driven by at least two factors. First, domains differ in how much the next item depends on stable long-term preference versus local sequential intent. Our recent-dominance analysis shows that Google Reviews and Books are less dominated by recent interactions, which is consistent with the strong gains of BLUE in these domains, since textual profiles can compactly summarize stable user interests. Second, domains also differ in how much success depends on fine-grained item distinctions. In Electronics, despite similarly low recent-dominance, BLUE is less competitive than classical sequential models, suggesting that the challenge is not short-term intent alone, but the need to preserve brand-, model-, and compatibility-level signals during embedding adaptation. By contrast, Clothing and Sports lie in between: BLUE still improves over other profile-based baselines, but sequential models remain competitive because recent transition patterns are relatively more important

\section{In-Domain Sequential Recommendation Performance Analysis}
\label{indomain_analysis}
To better understand domain-wise differences in in-domain sequential recommendation performance (Section~\ref{in-domain}), we define a \textit{recent-dominance score}. Specifically, we use \textit{Qwen3-Emb-0.6B} to embed the target item title and the titles of the most recent 10 user interactions, and compute their mean cosine similarity, denoted as $s_{\text{recent}}$. We similarly compute the mean similarity between the target item and all previous user interactions as $s_{\text{long}}$. The recent-dominance score is then defined as $s_{\text{recent}} - s_{\text{long}}$, which quantifies the extent to which the next item is driven by recent behavior rather than long-term preference. We observe that domain-wise performance differences are primarily influenced by two factors. First, domains vary in the relative importance of long-term preference versus short-term sequential intent. Our recent-dominance analysis shows that Google Reviews and Books exhibit lower reliance on recent interactions, suggesting that user behavior in these domains is more driven by stable preferences. This aligns with the strong performance of BLUE, as textual profiles can effectively summarize long-term user interests. Second, domains differ in the degree to which success depends on fine-grained item distinctions. In Electronics, despite relatively low recent-dominance, BLUE is less competitive than classical sequential models. This suggests that the main challenge is not short-term intent modeling, but preserving fine-grained signals such as brand, model, and compatibility during embedding adaptation. In contrast, Clothing and Sports lie between these extremes: while BLUE consistently improves over other profile-based methods, sequential models remain competitive due to the stronger role of recent transition patterns. Overall, these results indicate that the effectiveness of profile-based methods depends not only on temporal dynamics but also on the granularity of item-level distinctions required by the domain.

\section{Implementation Details}

In our implementation, both the maximum input and output generation lengths are constrained to 2048 tokens. We train the model for one epoch on the Amazon Clothing dataset. For Group Relative Policy Optimization (GRPO) training, we set the learning rate to 1e-6 and the batch size to 64. During this phase, we employ a sampling temperature of 1.0 with 8 rollouts per user, and sample 128 negative items per positive item to provide abundant negative feedback for user profile generation. For the KL divergence, we set the coefficient to 0.01 to keep the profiler model generating fluent user profiles. For the embedding model training via supervised contrastive learning, we use a batch size of 512 and a learning rate of 1e-6, restricting training to a single epoch to achieve domain adaptation while preserving generalization capabilities. The temperature for both the InfoNCE-style reward and the supervised contrastive loss is set to 0.07. Finally, during inference, the temperature is set to 0.7 for profile generation, whereas the decoding temperature is strictly set to 0 for the text-space next-item prediction task to eliminate randomness during evaluation.

\section{Computation Resources}

All the training experiments in this paper were conducted on we use the 64 TPU v6e (Trillium) chips with a topology of 8x8.

\section{Scientific Artifacts}
\ourmethod{} is built with the help of many existing scientific artifacts, including Jax~\citep{jax2018github}, and Simply~\citep{Liang2025Simply}.

\section{Discussion on Potential Data Contamination}
\label{app:data_contamination}

Large language models and embedding models are commonly pretrained on large-scale public web corpora. Since Amazon Reviews and Google Local Reviews are public datasets, we cannot rule out the possibility that some item descriptions, reviews, or related textual content were included in the pre-training data of the backbone models used in this work. This creates a potential data contamination concern: a pretrained model may have seen part of the benchmark text before evaluation, which could inflate absolute performance, especially in tasks that rely heavily on textual semantics.

Following prior work on zero-shot cross-domain sequential recommendation~\citep{li2025llm}, we use the term ``zero-shot'' to describe the evaluation protocol rather than the pre-training history of the backbone models. Specifically, zero-shot in our setting means that the model is trained on the source domain and evaluated on target domains without target-domain fine-tuning or task-specific training. It does not imply that the pretrained language or embedding models had no possible exposure to public benchmark data during pre-training.

This limitation affects the interpretation of absolute performance numbers. In particular, if the backbone models have memorized public item text or reviews, the reported results may overestimate performance compared with a fully unseen private benchmark. However, our main conclusions are based on relative comparisons under the same pretrained backbones, datasets, and evaluation protocol. All profile-based baselines and BLUE use the same profiler and embedding models, so possible pre-training exposure is shared across methods. Therefore, while potential contamination may affect the absolute scale of performance, it is less likely to fully explain the relative gains of BLUE over baselines.

To further reduce this concern, future work should evaluate BLUE on temporally held-out data released after the backbone models' pre-training cutoff, newly collected private interaction logs, or benchmarks with controlled data provenance. Such evaluation would provide a stricter test of whether optimized textual profiles generalize beyond public benchmark exposure.

\section{Ethical Considerations}

\paragraph{Privacy and user data protection.}
\ourmethod{} is trained from user interaction histories, which may contain sensitive signals about personal interests, habits, purchases, or visited locations. Although the model generates concise textual profiles instead of exposing raw histories directly, these profiles can still reveal private information if they are stored, inspected, or shared without proper safeguards. In addition, the learned embeddings and generated profiles may retain behavioral patterns that enable user inference. Any deployment should therefore apply strong privacy protections, including data minimization, secure storage, controlled access, anonymization where possible, and clear user consent and data retention policies.

\paragraph{Bias, fairness, and harmful inferences.}
User histories often reflect existing social and platform biases, and a profiling system trained on such data may reproduce or amplify these patterns. This can lead to unfair personalization outcomes, overgeneralized user summaries, or the implicit inference of sensitive attributes that users never explicitly provided. Because \ourmethod{} optimizes profiles for downstream utility, it may also learn predictive shortcuts that improve recommendation performance while encoding undesirable correlations. Careful evaluation is therefore needed to examine bias in both generated profiles and downstream predictions, especially across different user groups and behavioral patterns.

\paragraph{Transparency, user control, and responsible use.}
A key motivation of \ourmethod{} is that textual profiles are more interpretable than latent embeddings, but interpretability alone does not guarantee correctness. Generated profiles may still omit important context, overstate weak preferences, or reflect temporary behaviors as if they were stable traits. In downstream personalization, such errors can mislead users or produce undesirable system behavior. For this reason, profile-based systems should treat generated profiles as editable summaries rather than authoritative descriptions. When possible, users should be allowed to inspect, revise, or reset their profiles, and deployments should avoid manipulative or high-stakes uses without additional safeguards and human oversight.

\section{Limitations}

Our work has several limitations. First, the current study focuses on sequential recommendation and personalized question answering built from user interaction histories in Amazon Reviews and Google Local Reviews, so it does not yet test broader personalization settings such as long-horizon dialogue, agent memory, or multimodal user behavior. Second, the reward design is still relatively simple: the text-space signal is based on next-item multiple-choice prediction, and the embedding-space signal is defined by a fixed contrastive objective, which does not directly optimize other desirable properties such as faithfulness, calibration, robustness to noisy histories, or long-term user satisfaction. Because the model is optimized for downstream reward, it may also under-represent subtle but relevant user traits that are less useful for the current objectives. Third, \ourmethod{} adds extra training and inference cost compared with directly encoding raw user history, since it requires profile generation, multiple rollouts during reinforcement learning, and additional reward computation. Although the method shows a good quality-efficiency trade-off in our experiments, this extra cost may still limit deployment in high-throughput settings.

\section{Case Study of Training Dynamics}
To better understand how \ourmethod{} changes profile generation during reinforcement learning, we present two qualitative examples that track the same user's profile across training steps. These examples show that the profiler does not simply become more fluent over time. Instead, it gradually learns to remove generic language, reduce unsupported speculation, and retain behavior patterns that are more useful for downstream personalization.

\paragraph{Amazon Sports Example.}
Tables~\ref{tab:profile_evolving} and \ref{tab:profile_evolving2} show the training dynamics for a user whose history is dominated by cycling-related purchases, including helmets, phone mounts, mirrors, tires, saddle bags, and other bike accessories, with a smaller number of camping and pet-related items. At \textit{0 step}, the generated profile is broad and somewhat generic. It identifies high-level themes such as comfort, durability, safety, and practicality, but it also contains speculative statements about what the user may buy next, such as a reflective vest, a cycling pump, or even a new bike, which are not well grounded in the observed history. The summary is fluent, but it behaves more like a generic recommendation narrative than a compact user profile. After \textit{200 steps}, the profile begins to focus more clearly on cycling gear and safety-related accessories, but it still introduces noisy or weakly supported details, such as an emphasis on child riders and repeated discussion of size and fit that is not central to the overall pattern. By \textit{400 steps}, the profile becomes more behavior-centered and more selective. It consistently highlights cycling gear, safety accessories, portability, and durable equipment, while reducing unsupported next-item speculation. Although the profile still contains some unnecessary wording, it better reflects the actual structure of the history: cycling is the dominant theme, camping items are secondary, and the summary is more aligned with the user's repeated high-rated purchases. This example suggests that training helps the profiler move from generic and overly predictive descriptions toward more grounded summaries that isolate the stable preference signals most useful for retrieval and downstream reasoning.

\paragraph{Google Local Reviews Example.}
Tables~\ref{tab:profile_evolving3}, \ref{tab:profile_evolving4}, and \ref{tab:profile_evolving5} present a second example from Google Local Reviews. This user has a broad activity history that includes parks, beaches, churches, retail stores, grocery stores, bakeries, restaurants, airports, and other local businesses, with many visits concentrated in Queens and other areas of New York. At \textit{0 step}, the profile captures some broad regularities, such as interest in outdoor places, food, and retail, but it is still quite diffuse and contains several weak abstractions, such as a vague preference for ``welcoming atmosphere'' and ``ample parking,'' which are not directly supported by the raw history. The summary also mixes many categories together without clearly separating recurring behavioral patterns from incidental visits. At \textit{200} and \textit{400 steps}, the profile becomes more stable in its emphasis on outdoor recreation, casual retail, familiar chains, and food-related locations, but it still includes noisy details such as references to ``baseball'' or overly generic claims about atmosphere and convenience. By \textit{600} and \textit{800 steps}, the profile better organizes the user's behavior into a few recurring themes: outdoor and park-related activities, casual dining and shopping, familiarity with common retail chains, and repeated movement across local community-oriented places in New York. While the later profiles are still somewhat verbose, they are more grounded in the actual visit history and less reliant on abstract lifestyle claims. This progression indicates that \ourmethod{} gradually learns to compress diverse location histories into a smaller set of reusable preference factors that are more likely to support both recommendation and language-based personalization.

\paragraph{Overall Observation.}
Across both examples, the main change over training is not only improved fluency, but improved selectivity and grounding. Early-stage profiles tend to over-explain, speculate about future actions, or introduce attributes that are only loosely supported by the history. As training proceeds, the model becomes better at identifying repeated behavioral signals and summarizing them in a form that is more compact, more faithful to the observed data, and more useful for downstream tasks. These examples provide qualitative evidence for our main claim: the reward signals in \ourmethod{} help the profiler move away from generic natural-language summaries and toward user profiles that better preserve behaviorally useful preference information.

\begin{table}[t]
   % \vspace{-0.3cm}
   \caption{User profile for user `AFFCQMK6WPS6MO6ANTWMCNZNU56Q' under Amazon Sports dataset during the training process. }
   \centering
   \small
   \begin{adjustbox}{max width=1\linewidth}
   \begin{tabular}{l|p{12cm}}
   \toprule[1.2pt]
   \textsc{Raw User History} & [Paracord Planet - Genuine Type III 550 Paracord Nylon Colors Multiple Sizes – 550 LB Tensile Strength | 4.0], [B\&W International Bike Sack - Bike Sack (96250/N) | 3.0], [White Lightning Chain Johnny | 5.0], [Wallfire Bike Phone Mount Bag Bicycle Frame Bike Handlebar Bags with Waterproof Touch Screen Phone Case Bicycle Phone Holder Storage Bag\_Black Red | 5.0], [Continental Gator Hardshell Folding Bike Tire - Puncture Protection Replacement Road/Commuter Tire (23c, 25c, 28c, 32c) | 5.0], [[Bone] Bike Phone Mount Universal 2 in 1 Phone Holder for Bike, Charger Mount Stem, Bike Accessories for iPhone 13 12 11 Pro Max Samsung S20 S21, Fit 4.7"-7.2" Phones (Bike Tie Pro Pack 2-Black) | 5.0], [Sprintech Road Drop Bar Rearview Bike Mirror - Cycling Safety Mirror - Single For Left Side Dropbar | 3.0], [Zacro Adult Bike Helmet Lightweight - Bike Helmet for Men Women Comfort with Pads\&Visor, Certified Bicycle Helmet for Adults Youth Mountain Road Biker | 4.0], [Everest Cooler Lunch Bag, Red, One Size | 4.0], [BV Bicycle Strap-On Saddle Bag with perfect Size I With reflective for a Safety ride I Seat Bag, Cycling Bag - Bike Bag for all our essentials, bike bags for bicycles, bicycle bag, bike seat bag | 5.0], [Schwinn Thrasher Adult Lightweight Bike Helmet, Dial Fit Adjustment, LED and Non-Lighted Options, Suggested Fit 58-62 Cm | 2.0], [POWERLIX Ultralight Sleeping Pad for Camping with Inflating Bag, Carry Bag, Repair Kit – Compact Lightweight Camping Mat, Outdoor Backpacking Hiking Traveling Airpad Camping Air Mattress | 5.0], [Pack-n-Wipe Disposable Compressed Towels - Absorbent \& Durable Cleaning Rags - Compact Cleaning Cloth Travel Essentials - Soft Baby Wipes \& Baby Towel (Pack-N-Wipe (100 Towels \& 4 Tubes)) | 5.0], [Enbrighten LED Camping Lantern, Battery Powered, Carabiner Handle, Hiking Gear, Emergency Light, Tent Light, Lantern Flashlight for Hurricane, Emergency, Survival Kits, Fishing, Home and More | 5.0], [Slant Board, Calf Stretcher Ankle and Foot Incline Board for Stretching Tight Calves or Plantar Fasciitis Adjustable 4 Level (330 lb Capacity), Strength Training Equipment Leg Machines | 3.0], [Pawtitas Pet Adjustable Solid Color Step in Puppy/Dog Harness 6 feet Matching Collar and Harness Sold Separately Small Grey/Blue/Green Multicolor | 4.0] \\
   % \hline
   % \textsc{0 step user profile} &  \\
   % \hline
   % \textsc{200 step user profile} & \\ 
   % \hline
   % \textsc{400 step user profile} & \\
   % \hline
   \midrule[0.75pt]
    \textsc{0 Step User Profile} & Okay, based on the user history, here’s a summary of their likely preferences and a prediction for their next purchase:

The user exhibits a strong preference for **comfort, durability, and safety** in their cycling gear. They consistently rate products highly (4.0-5.0) across a wide range of categories, particularly those related to cycling accessories and protection.

Specifically, they seem to prioritize **quality and reliability**. The frequent purchase of high-quality items like Paracord Planet, Wallfire Bike Phone Mount, and Continental Gator Hardshell Tire demonstrates a commitment to durable equipment. They also appreciate well-designed and functional accessories like bike helmets and reflective cycling gear.

Furthermore, there’s a clear interest in **practicality and versatility**. The inclusion of a cycling mirror, a backpack, and a step in harness, along with items like a sleeping pad and portable lantern, suggests an individual who values preparedness and ease of use for their cycling activities.

Considering this, the user is likely looking for solutions that prioritize both performance and safety. Given their existing purchases, a likely next step is a **replacement for their existing cycling helmet or a new bike accessory**, perhaps a reflective vest or a high-quality cycling pump. They're also open to expanding their gear collection, suggesting a desire for more robust and reliable cycling solutions.\\
   \bottomrule[1.2pt]
   \end{tabular}
   \end{adjustbox}
   \vspace{-0.3cm}
   \label{tab:profile_evolving}
\end{table}

\begin{table}[t]
   % \vspace{-0.3cm}
   \caption{User profile for user `AFFCQMK6WPS6MO6ANTWMCNZNU56Q' under Amazon Sports dataset during the training process. (Continued) }
   \centering
   \small
   \begin{adjustbox}{max width=1\linewidth}
   \begin{tabular}{l|p{12cm}}
   \toprule[1.2pt]
   \textsc{200 Step User Profile} & The user displays a clear interest in comfortable, durable, and well-suited cycling gear, particularly for both adult and child riders. They favor lighter-weight, compact, and reliable equipment with high quality ratings (5.0+). They demonstrate an appreciation for adjustable accessories like bike helmets and straps, with a focus on safety and comfort ("Certified Bicycle Helmet for Adults," "Adjustable Fit").

The user has a clear preference for size and fit (e.g., 4.0, 5.0, 3.0), indicating a comfortable and secure fit for their bike. They also demonstrate an interest in accessories like phone holders and reflective gear ("reflective for a Safety ride I"). They favor brands and styles with waterproof features ("Waterproof Touch Screen Phone Case") and are receptive to items with a durable and robust build (“Heavy Duty”).

The user is actively buying a “Paracord Planet” (Genuine Type III 550 Paracord Nylon Colors Multiple Sizes – 550 LB Tensile Strength) and a “Zacro Adult Bike Helmet” (4.0). Based on this history, the user is likely looking for a comfortable and reliable cycling helmet that prioritizes safety and durability, and they are open to purchasing a suitable backpack and cycling accessories. They're likely ready to purchase a new bike, perhaps a new helmet, or a new bike bag. \\
   \hline
   \textsc{400 step user profile} & The user displays a clear interest in comfortable, durable cycling gear, safety accessories, and portable hydration solutions. They are highly responsive to high-quality, well-rated items with a focus on comfort and protection ("Genuine Type III Paracord Planet", “Wallfire Bike Phone Mount”). The user demonstrates an appreciation for compact and versatile gear, including bike accessories (bike sacks, phone mounts), and is responsive to various sizing and fit options ("Continental Gator Hardshell Folding Bike Tire", "[Bone] Bike Phone Mount").

The user is also responsive to safety and portability ("Sprintech Road Drop Bar Rearview Bike Mirror"), and has a significant interest in cycling safety (helmet, reflective gear). They favor items with adjustable straps and fit, and demonstrate an understanding of appropriate sizes (“Schwinn Thrasher Adult Lightweight Bike Helmet”). They're also responsive to smaller, adjustable products and are very comfortable with soft, comfortable materials (Paracord Planet).

The user is seeking a product that is durable, comfortable, and offers a secure fit for cycling and outdoor adventures. Based on this history, the user is likely to be looking for a comfortable, safe, and reliable cycling accessory, or a compact, portable camping solution. They are very responsive to positive ratings and are likely to purchase a product that aligns with their comfort and safety needs. Next product purchase will likely be a comfortable, well-fitting bike accessory or a portable camping item. \\
   % \hline
   \bottomrule[1.2pt]
   \end{tabular}
   \end{adjustbox}
   \vspace{-0.3cm}
   \label{tab:profile_evolving2}
\end{table}

\begin{table}[t]
   % \vspace{-0.3cm}
   \caption{User profile for user `102100435192911638466' under Google Local Review dataset during the training process. }
   \centering
   \small
   \begin{adjustbox}{max width=1\linewidth}
   \begin{tabular}{l|p{12cm}}
   \toprule[1.2pt]
   \textsc{Raw User History} & [The Home Depot | The Home Depot, 50-10 Northern Blvd, Queens, NY 11101 | 5], [NY Pizza Suprema | NY Pizza Suprema, 413 8th Ave, New York, NY 10001 | 5], [FOOD BAZAAR | FOOD BAZAAR, 35-60 Junction Blvd, Queens, NY 11368 | 5], [Wildwood State Park | Wildwood State Park, 790 Hulse Landing Rd, Wading River, NY 11792 | 5], [Raphael | Raphael, 39390 NY-25, Peconic, NY 11958 | 5], [Bailie Beach Park | Bailie Beach Park, 1901 Bailie Beach Rd, Mattituck, NY 11952 | 5], [Grand Army Plaza | Grand Army Plaza, NY 10019 | 5], [Mattituck Park District | Mattituck Park District, 5799-5003 Luthers Rd, Mattituck, NY 11952 | 5], [Sueño Americano | Sueño Americano, 35-16 Junction Blvd, Queens, NY 11372 | 4], [Roosevelt Field | Roosevelt Field, 630 Old Country Rd, Garden City, NY 11530 | 5], [Mamajuana Cafe Queens | Mamajuana Cafe Queens, 33-15 56th St, Queens, NY 11377, United States | 5], [Sam's Club | Sam's Club, 2950 Horseblock Road, Medford, NY 11763 | 5], [FOOD BAZAAR | FOOD BAZAAR, 34-20 Junction Blvd, Queens, NY 11372 | 5], [St Nicholas Roman Catholic Church | St Nicholas Roman Catholic Church, 26 Olive St, Brooklyn, NY 11211 | 5], [Daniel’s Bait \& Tackle Shop | Daniel’s Bait \& Tackle Shop, 1184, 3348 103rd St, Corona, NY 11368 | 5], [St. Charles Borromeo Church, Brooklyn Heights | St. Charles Borromeo Church, Brooklyn Heights, 19 Sidney Pl, Brooklyn, NY 11201 | 5], [The Home Depot | The Home Depot, 73-01 25th Ave, Queens, NY 11370 | 5], [North Sea Beach | North Sea Beach, Southampton, NY | 5], [LaRoma Pizzeria \& Restaurant of Rome, NY | LaRoma Pizzeria \& Restaurant of Rome, NY, 600 Floyd Ave, Rome, NY 13440 | 5], [Walmart Supercenter | Walmart Supercenter, 5815 Rome-Taberg Rd, Rome, NY 13440 | 5], [LaGuardia Airport | LaGuardia Airport (LGA), Queens, NY 11371 | 5], [Northern Playground | Northern Playground, Northern Blvd, Jackson Heights, NY 11372 | 5], [McDonald's | McDonald's, 37-19 Junction Blvd, Queens, NY 11368 | 5], [Sunken Meadow State Park | Sunken Meadow State Park, New York State Reference Rte 908K, Kings Park, NY 11754 | 5], [Croton Gorge Park | Croton Gorge Park, 35 Yorktown Rd, Croton-On-Hudson, NY 10520 | 5], [Bloomingdale's | Bloomingdale's, 630 Old Country Rd, Garden City, NY 11530 | 5], [McDonald's | McDonald's, 276 Service Rd, Amsterdam, NY 12010 | 5], [Christ Church, Oyster Bay | Christ Church, Oyster Bay, 61 E Main St, Oyster Bay, NY 11771 | 5], [COLLEGE POINT CENTER | COLLEGE POINT Center, 137-05 20th Ave, College Point, NY 11356 | 5], [Subway | Subway, 204 E Thomas St, Rome, NY 13440 | 5], [Classic Car Wash | Classic Car Wash, 7404 Northern Blvd, Jackson Heights, NY 11372 | 5], [Schmitt's Farm Stand on Sound | Schmitt's Farm Stand on Sound, 3355 Sound Ave, Riverhead, NY 11901 | 5], [Macy's | Macy's, 9001 Queens Blvd, Queens, NY 11373 | 5], [McDonald's | McDonald's, 3789 NY-13, Pulaski, NY 13142 | 5], [Seba-Seba Bakery | Seba-Seba Bakery, 83-03 Northern Blvd, Flushing, NY 11372 | 5], [Fat Nancy's Tackle Shop | Fat Nancy's Tackle Shop, 3750 NY-13, Pulaski, NY 13142 | 5], [Las Americas Bakery | Las Americas Bakery, 9304 A 37th Ave, Queens, NY 11372 | 5], [E-ZPass NY | E-ZPass NY, 19-02 Whitestone Expy \#101, Queens, NY 11357 | 5], [Gorman Playground | Gorman Playground, 25th Ave. \& 84th St, East Elmhurst, NY 11369 | 5], [McDonald's | McDonald's, 51-35 Northern Blvd, Queens, NY 11377 | 5] \\
   % \hline
   \bottomrule[1.2pt]
   \end{tabular}
   \end{adjustbox}
   \vspace{-0.3cm}
   \label{tab:profile_evolving3}
\end{table}

\begin{table}[t]
   % \vspace{-0.3cm}
   \caption{User profile for user `102100435192911638466' under Google Local Review dataset during the training process. (Continued) }
   \centering
   \small
   \begin{adjustbox}{max width=1\linewidth}
   \begin{tabular}{l|p{12cm}}
   \toprule[1.2pt]
   \textsc{0 Step User Profile} & Okay, here’s a summary of the user’s preferences, designed to inform a prediction of their next location choice, within 200 words:

The user exhibits a strong preference for outdoor recreation and a broad range of retail options, primarily centered around home improvement and food. They frequently visit locations offering a variety of products, ranging from hardware stores (Home Depot, Sam’s Club, Sears) to specialty food stores (FOOD BAZAAR, LaRoma Pizzeria, Las Americas Bakery).

The user demonstrates a consistent interest in Queens and surrounding areas, with frequent visits to locations like Wildwood State Park, Roosevelt Field, and North Sea Beach. Their history reveals a preference for locations with a welcoming atmosphere, indicating a desire for a pleasant experience.

Furthermore, the user’s frequent visits to locations with a focus on “hardware” and “food” suggests a need for supplies and sustenance. They also show a preference for locations with a strong presence of “family” – evidenced by repeated visits to places like Bailey Beach Park, and a familiarity with locations like St. Charles Borromeo Church.

Based on this, the user is likely looking for a destination that combines outdoor activities, a variety of goods, and a comfortable environment. They're likely to appreciate a location with ample parking and a welcoming ambiance. Given their established preferences, a suggestion of a local park, a large grocery store, or a popular restaurant in the area would be a logical next step. \\
\hline
\textsc{200 Steps User Profile} & The user exhibits a strong interest in outdoor recreational activities, particularly with a focus on “beach” and “park” locations. They consistently favor locations with a “5” rating (indicating high satisfaction), suggesting a preference for accessible and pleasant environments. Their history reveals a clear interest in “wildlife” (Wildwood State Park, Northern Playground) and “baseball” (Roosevelt Field, Baseball).

The user displays a preference for familiar brands and stores – “Sam’s Club,” “Bailie Beach Park,” “The Home Depot,” “McDonald’s,” “Walmart Supercenter.” They also appreciate a casual, local atmosphere, as evidenced by their preference for “Raphael” and “Mamaunana Cafe Queens.”

They’re drawn to locations with a relaxed pace and are comfortable with a comfortable “5” rating. Based on this, the user is likely looking for a nearby park, beach, or recreational area, and is comfortable with a quick and easy experience. They’re most likely to enjoy a casual, familiar, and enjoyable outing – suggesting a desire to visit a familiar establishment. The user is likely seeking a location with a friendly atmosphere and a comfortable environment. \\
\hline
\textsc{400 Steps User Profile} & The user displays a clear interest in outdoor recreational activities, particularly beach, park, and sports-related locations ("Wildwood State Park", "North Sea Beach", "St. Charles Borromeo Church", "Northern Playground"). They demonstrate a comfortable familiarity with various retail chains ("Sam's Club", "The Home Depot", "Walmart Supercenter", "Bloomingdale's") and are responsive to locations with a focus on outdoor activities and local community ("LaRoma Pizzeria \& Restaurant of Rome", "Sunken Meadow State Park").

The user exhibits a preference for a casual, family-friendly atmosphere (“Bailie Beach Park”, “LaRoma Pizzeria \& Restaurant of Rome”) and are comfortable with a range of food options (“FOOD BAZAAR”, “LaGuardia Airport”). They favor locations with a focus on the outdoors and are responsive to the proximity of parks and recreational areas.

The user is a repeat user, showing a consistent pattern of frequent visits to various locations. They are comfortable with a range of retail and food options and are responsive to the proximity of outdoor recreational areas. The user is looking for a comfortable, casual, and convenient location that offers a relaxed atmosphere and a variety of activities. The user is likely to respond to a location with amenities and a welcoming atmosphere. \\
   % \hline
   \bottomrule[1.2pt]
   \end{tabular}
   \end{adjustbox}
   \vspace{-0.3cm}
   \label{tab:profile_evolving4}
\end{table}

\begin{table}[t]
   % \vspace{-0.3cm}
   \caption{User profile for user `102100435192911638466' under Google Local Review dataset during the training process. (Continued) }
   \centering
   \small
   \begin{adjustbox}{max width=1\linewidth}
   \begin{tabular}{l|p{12cm}}
   \toprule[1.2pt]
\textsc{600 Steps User Profile} & The user displays a clear interest in outdoor recreational activities, specifically related to a variety of locations within New York State and surrounding areas. The user displays a preference for casual, family-friendly experiences with a focus on coastal and park-like environments ("Wildwood State Park", "North Sea Beach", "Northern Playground", "SUNKEN MEADOW STATE PARK", "Croton Gorge Park", "Bloomingdale's").

They favor locations with a strong emphasis on retail (Home Depot, Sam’s Club, Macy’s) and dining (LaRoma Pizzeria \& Restaurant of Rome, McDonald’s, Classic Car Wash, Las Americas Bakery). The user demonstrates an affinity for casual, familiar establishments with a focus on food and drink ("FOOD BAZAAR", "LaGuardia Airport", "Schmitt’s Farm Stand on Sound").

They seem comfortable with a variety of store types – from large retailers to smaller, local businesses. They're responsive to locations with a strong community feel ("St. Charles Borromeo Church", "Northern Playground"). They're also responsive to locations with easy access to transportation (E-ZPass NY, Subway).

They're comfortable with a range of price points, from budget-friendly to mid-range. They’re likely looking for comfortable, convenient access to a variety of amenities and are comfortable with a range of food options. They’re comfortable with a variety of store types.” \\
\hline
\textsc{800 Steps User Profile} & The user displays a strong interest in outdoor recreation, particularly beach and park activities ("Wildwood State Park", "North Sea Beach", "Sueno Americano", "St. Charles Borromeo Church", "The Home Depot", "Northern Playground", "LaRoma Pizzeria \& Restaurant of Rome"). They favor casual, family-friendly establishments ("FOOD BAZAAR", "McDonald's", "Classic Car Wash", "Schmitt's Farm Stand on Sound"). They demonstrate a preference for casual dining with American-style food ("Food Bazaars", "McDonald's"). They also appreciate a comfortable and accessible location (“The Home Depot”, “Seba-Seba Bakery”).

Based on this history, the user is responsive to locations with easy access to outdoor spaces and a comfortable atmosphere. They’re comfortable with a wide range of food options, including casual American fare. They’re likely seeking a relaxed and enjoyable experience, and are responsive to locations with a focus on community and local businesses. They’re comfortable with a variety of stores, and appreciate a dependable service. They’re interested in comfortable, casual environments. Therefore, the user is seeking a location with a welcoming, comfortable experience, with an emphasis on recreational activities and food.""" \\
   % \hline
   \bottomrule[1.2pt]
   \end{tabular}
   \end{adjustbox}
   \vspace{-0.3cm}
   \label{tab:profile_evolving5}
\end{table}

\section{Prompt Details}
We present the prompt template for user profile generation in Table~\ref{profile_gen} and user history in Table~\ref{history_format}

\begin{tcolorbox}
{\fontfamily{cmss}\selectfont
You are an expert recommendation system. The user history is provided in the format of [product\_name | rating], where rating=1 is the lowest and rating=5 is the highest. Your task is to summarize the user interest, preferences, and that can be directly used to predict what product the user would likely to purchase next. \\
User History: \\
\{user\_history\} \\
Summarize the preferences of the user that can be used to predict the user's next behavior within 200 words.
}
\end{tcolorbox}
\noindent\begin{minipage}{1.0\textwidth}
\captionof{table}{Prompt template for user profile generation in \ourmethod{}, RLAIF, RLPF, and LangPTune for fair comparison.}\label{profile_gen}
\end{minipage}

\begin{tcolorbox}
{\fontfamily{cmss}\selectfont
\% \textit{User History Format in Amazon Review Data}\\{}
[\{ product\_name\} | \{rating\}] \\

\% \textit{User History Format in Google Local Review Data}\\{}
[\{location\_name\} | \{address\} | \{rating\}]
}
\end{tcolorbox}
\noindent\begin{minipage}{1.0\textwidth}
\captionof{table}{User history format.}\label{history_format}
\end{minipage}

\end{document}